\documentclass[12pt]{article}
\usepackage{epsfig,textpos,braket,url}
\usepackage{lineno}

\newcommand\diag{\mathop{\mathrm{diag}}}
\newcommand\real{\mathop{\mathrm{real}}}

\newcommand\tV{\tilde V}

\newcommand\ho{\hat\omega}
\newcommand\opi{\omega_{\pi}}
\newcommand\hC{\hat C}

\newcommand\Do{\Delta\omega}
\newcommand\Dk{\Delta\kappa}
\newcommand\dt{\Delta t}
\newcommand\oo{\omega_{12}}
\begin{document}
\title{The transient behavior of superconducting\\ multi-cell accelerating cavities.}
\author{Volker Ziemann\\ Thomas Jefferson National Accelerator Facility\\
  Newport News Va, USA}
\date{February 26, 2026}
\maketitle
\begin{abstract}\noindent
  We employ an equivalent-circuit model of a multi-cell cavity to explore its time-dependent
  behavior in order to understand differences between the multi-cell model and the commonly-used
  model of a single-cell resonator. Furthermore, we address tolerances that arise from
  manufacturing imperfections.
\end{abstract}
%
%
\section{Introduction}
In most simulations of the transient behavior of multi-cell superconducting
acceleration structures the cavity is represented by a single resonator
that consists of a discrete resistor $R$, inductor $L$, and capacitance
$C$~\cite{SCHILCHER,SYSID}. In reality, however, the individual cells
that constitute the cavity should be treated as individual but coupled
resonators, each having its specific resonance-frequency detuning,
$Q$-value, and shunt impedance~$R$. Only for the analysis of the supported
eigenmodes in multi-cell cavities, the cells are considered as coupled
resonators~\cite{PKH}. This analysis is mostly static though some 
time-dependent effects are mentioned in~\cite{LIEPE}. In this report,
we present a comprehensive analysis of multi-cell cavities and discuss 
the impact of cavity-fabrication imperfections on their time-dependent behavior.
\par
\begin{figure}[tb]
\begin{center}
\includegraphics[width=0.99\textwidth]{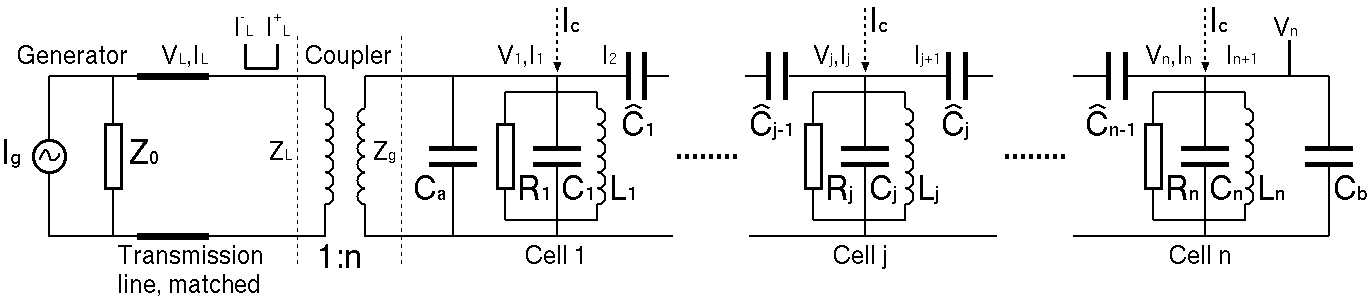}
\end{center}
\caption{\label{fig:mcc}Multi-cell cavity based on parallel-circuit RLC resonators,
  coupled via capacitors $\hC$, and driven by a generator with input impedance $Z_0$.}
\end{figure}
Figure~\ref{fig:mcc} illustrates our model. The generator with internal
impedance $Z_0$ provides a current $I_g$ that is transported to the
cavity by a matched transmission line. The input power is fed to the
first cell with an input coupler that is modeled as a transformer with
winding ratio $n$. Each of the individual cells is represented as an
independent RLC circuit and adjacent cells are coupled via capacitors
$\hat C$. Additionally, the first and last cell experience the coupling
to the beam pipe, here modeled by the capacitors labeled $C_a$ and $C_b$.
We also indicate the beam current $I_c$ that passes through all cells and
excites fields inside the cells, though in this note we do not consider
this effect and focus on the behavior of the cavity without beam. In
particular, we address variations among cells and how they affect the
forward current $I^+_L$ and the backward current $I^-_L$ measured by a
directional coupler just upstream of the coupler as well as a field
probe connected to the last cell that measures the cavity voltage $V_n$.
Reference planes on either side of the coupler are depicted as dashed
lines. The voltages and currents are related by the transformer ratio
$V_1=nV_L$ and $I_1=I_L/n$ and consequently the impedances $Z_0=Z_L=V_L/I_L$
and $Z_g=V_1/I_1$ are related by $Z_g=n^2Z_L$.
\par
Before addressing the fully time-dependent problem, we briefly address the
steady state of the bare multi-cell cavity. This analysis follows the
discussion in~\cite{PKH} and helps us to determine values for the beam pipe
capacitances $C_a$ and $C_b$ as well as the supported eigenfrequencies and
corresponding eigenmodes in the cavity. In contrast to the discussion
in~\cite{PKH,LIEPE}, which is based on series-resonator circuits, we use
parallel circuits.
\section{Steady-state}
\label{sec:ss}
%
\begin{figure}[tb]
\begin{center}
\includegraphics[width=0.8\textwidth]{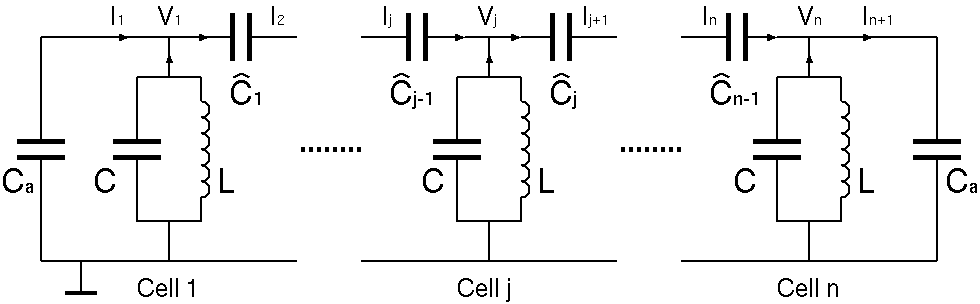}
\end{center}
\caption{\label{fig:mcb}Bare multi-cell cavity in which coupling to the generator
  and the resistors are omitted. Note, however, the capacitors $C_a$ and $C_b$ that
  represent coupling to the beam pipe.}
\end{figure}
In order to determine $C_a$ and $C_b$ we analyze the simplified model shown in
Figure~\ref{fig:mcb} where the generator and coupler are omitted. Moreover, all
capacitances between the cavities are assumed equal and that $C_a=C_b$ owing to
the forward-backward symmetry of the simplified model. We add a grounding symbol
to indicate the reference level for voltage measurements.
\par
We assume that the system oscillates with frequency $\omega$ such that we can use
the following impedances $Z_C=1/i\omega C$ and $Z_L=i\omega L$ to characterize
the capacitor and inductor, respectively. For the current balance at the first
node with voltage $V_1$ we thus find
\begin{equation}
  0=I_1-I_2+I_{LC}
  =i\omega C_a V_1 -i\omega \hC (V_2-V_1) + \left(i\omega C+\frac{1}{i\omega L}\right) V_1\ ,
\end{equation}
where $I_{LC}$ is the current that flows from ground to the node labeled $V_1$. Diving this
equation by $i\omega C$ and introducing the abbreviations $\kappa_a=C_a/C$ and $\ho^2=1/LC$
we arrive at
\begin{equation}\label{eq:c1}
  \left(1+\kappa+\kappa_a\right)V_1-\kappa V_2 = \frac{\ho^2}{\omega^2}V_1\ .
\end{equation}
Likewise, the current balance at cell~$j$ leads us to
\begin{equation}
  0=I_j-I_{j-1}+I_{LC}
  = i\omega\hC(V_j-V_{j-1}) - i\omega\hC(V_{j+1}-V_j) + \left(i\omega C+\frac{1}{i\omega L}\right) V_j\ .
\end{equation}
Dividing by $i\omega C$ and some rearranging then gives us
\begin{equation}\label{eq:cj}
  -\kappa V_{j-1}+(1+2\kappa)V_j-\kappa V_{j+1} =\frac{\ho^2}{\omega^2}V_j\ .
\end{equation}
Repeating the analysis for the right-most node labeled $V_n$ finally results in
\begin{equation}
0=i\omega\hC(V_n-V_{n-1})+ \left(i\omega C+\frac{1}{i\omega L}\right) V_j - i\omega C_a (0-V_n)
\end{equation}
and, after introducing abbreviations, leads to
\begin{equation}\label{eq:cn}
  -\kappa V_{n-1}+(1+\kappa+\kappa_a)V_n= \frac{\ho^2}{\omega^2}V_n\ .
\end{equation}
Jointly, equations~\ref{eq:c1},~\ref{eq:cj}, and~\ref{eq:cn} can be cast into the
following eigenvalue equation
\begin{equation}\label{eq:evs}
  \left(
    \begin{array}{ccccc}
      1+\kappa+\kappa_a & -\kappa & 0 & 0 & 0 \\
      -\kappa & 1+ 2\kappa & -\kappa & 0 & 0 \\
      0  &-\kappa & 1+ 2\kappa & -\kappa &  \\
      0 &0  &-\kappa & 1+ 2\kappa & -\kappa \\
      0  & 0  & 0  & -\kappa & 1+\kappa+\kappa_a
    \end{array}
  \right)
  \left(\begin{array}{c} V_1\\ V_2\\ V_3\\ V_4\\ V_5\end{array}\right)
  = \frac{\ho^2}{\omega^2}
  \left(\begin{array}{c} V_1\\ V_2\\ V_3\\ V_4\\ V_5\end{array}\right)
\end{equation}
where we assumed that $n=5$. Generalizing to other numbers of cells $n$ is trivial.
\par
We follow~\cite{PKH} and require the $\pi$-mode to produce alternating signs of
voltages in adjacent cells: $(V_1,V_2,V_3,\dots)=(1,-1,1,\dots)$. Imposing this
requirement on Equation~\ref{eq:evs} and considering the first two rows leads
to the conditions
\begin{equation}
  1+2\kappa+\kappa_a  = \frac{\ho^2}{\opi^2}
  \qquad\mathrm{and}\qquad
  -1-4\kappa = -\frac{\ho^2}{\opi^2}\ .
\end{equation}
Solving for $\kappa_a$ and $\ho^2/\opi^2$ leads to
\begin{equation}\label{eq:sol}
  \kappa_a=2\kappa
  \qquad\mathrm{and}\qquad
  \opi = \frac{\ho}{\sqrt{1+4\kappa}}\ .
\end{equation}
We need to point out that in our model the
$\pi$-mode frequency lies below the cell's resonance frequency $\ho$, which
is an artefact of using a parallel-resonator circuit. With a series-resonator
circuit the $\pi$-mode frequency is above $\ho$. 
\par
Now that we know what $\kappa_a$ should be, we insert it in Equation~\ref{eq:evs}
and find the following, now properly matched, eigenvalue equation
\begin{equation}\label{eq:evs2}
  \left(
    \begin{array}{ccccc}
      1+3\kappa & -\kappa & 0 & 0 & 0 \\
      -\kappa & 1+ 2\kappa & -\kappa & 0 & 0 \\
      0  &-\kappa & 1+ 2\kappa & -\kappa &  \\
      0 &0  &-\kappa & 1+ 2\kappa & -\kappa \\
      0  & 0  & 0  & -\kappa & 1+3\kappa
    \end{array}
  \right)
  \left(\begin{array}{c} V_1\\ V_2\\ V_3\\ V_4\\ V_5\end{array}\right)
  = \frac{\ho^2}{\omega^2}
  \left(\begin{array}{c} V_1\\ V_2\\ V_3\\ V_4\\ V_5\end{array}\right)
\end{equation}
that we inspect further to determine all eigenfrequencies and eigenmodes.
As mentioned before, the corresponding equations in~\cite{PKH} are
derived from a series-circuit model as opposed to the parallel-circuit model
employed here. Therefore, even though the structure of Equation~\ref{eq:evs2}
resembles the corresponding equation from~\cite{PKH}, the definition of
constants differs and therefore no one-to-one correspondence is possible.
\begin{figure}[tb]
\begin{center}
\includegraphics[width=0.9\textwidth]{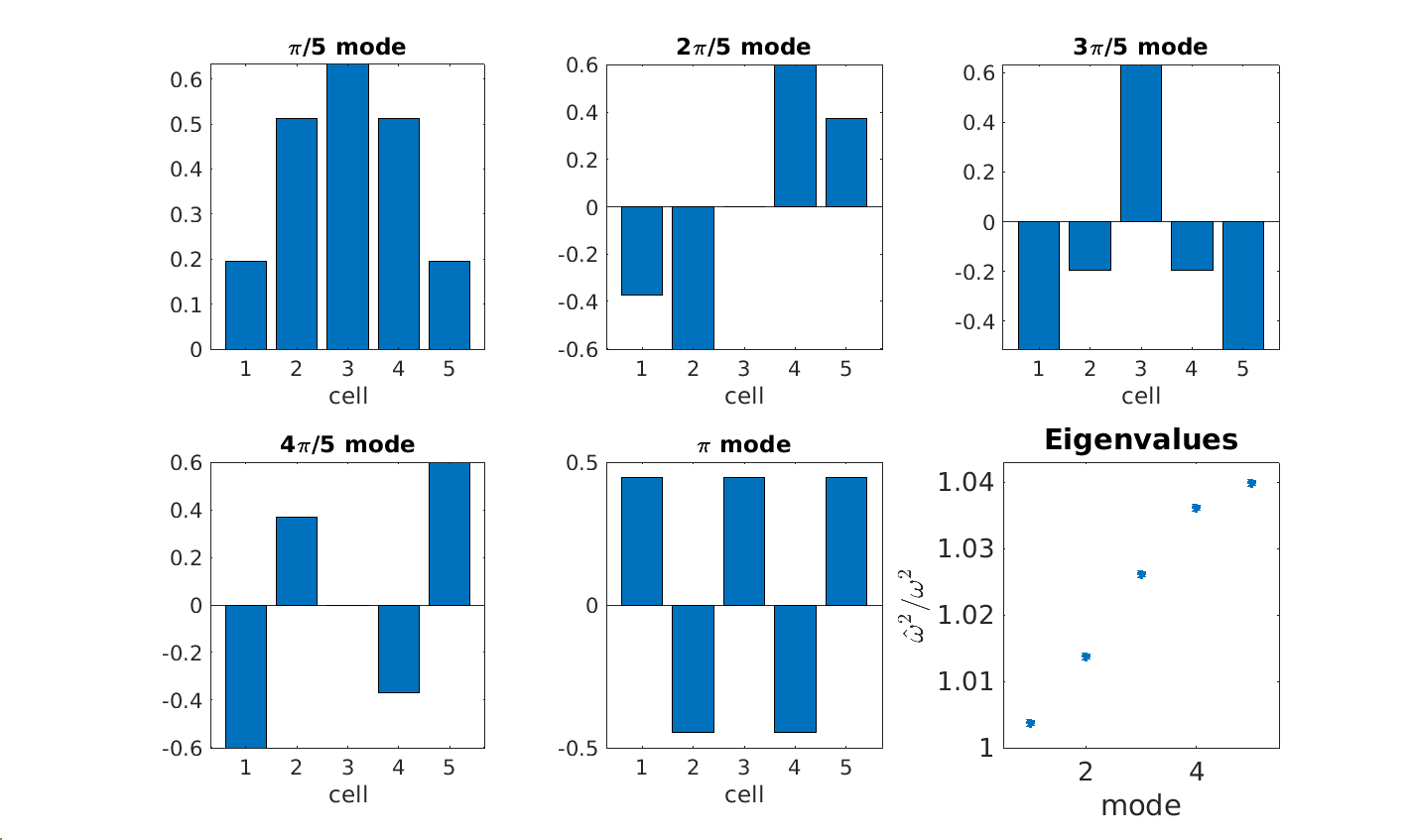}
\end{center}
\caption{\label{fig:evs}Modes and eigenvalues of a five-cell cavity with
  $\kappa=0.01$.}
\end{figure}
\par
As a matter of fact, the high degree of symmetry of the matrix in
Equation~\ref{eq:evs2} permits an analytical representation of the eigenvalues
and eigenvalues, which is used in~\cite{PKH}, but we do not duplicate those
results. Instead, we calculate the
corresponding quantities numerically using Matlab. Figure~\ref{fig:evs}
shows the five eigenmodes for $\kappa=0.01$ in the plots labeled $n\pi/5$-mode
and the corresponding eigenvalues in the bottom-right plot. Here mode~1
refers to the $\pi/5$-mode and mode~2 to the $2\pi/5$-mode. In particular,
we see that the largest eigenvalue for mode~5 corresponds to the $\pi$-mode
whose numerical value is $\hat\omega^2/\opi^2=1.04$, which agrees with
Equation~\ref{eq:sol}.
\par
If we excite the different modes with their respective frequencies, the combined
excitation pattern can be shaped to peak in different cells. This feature is
exploited when plasma-processing a cavity~\cite{PLASMA}. It allows us to locally
excite a plasma in a particular cell to clean it.
\par
\begin{figure}[tb]
\begin{center}
  \includegraphics[width=0.47\textwidth]{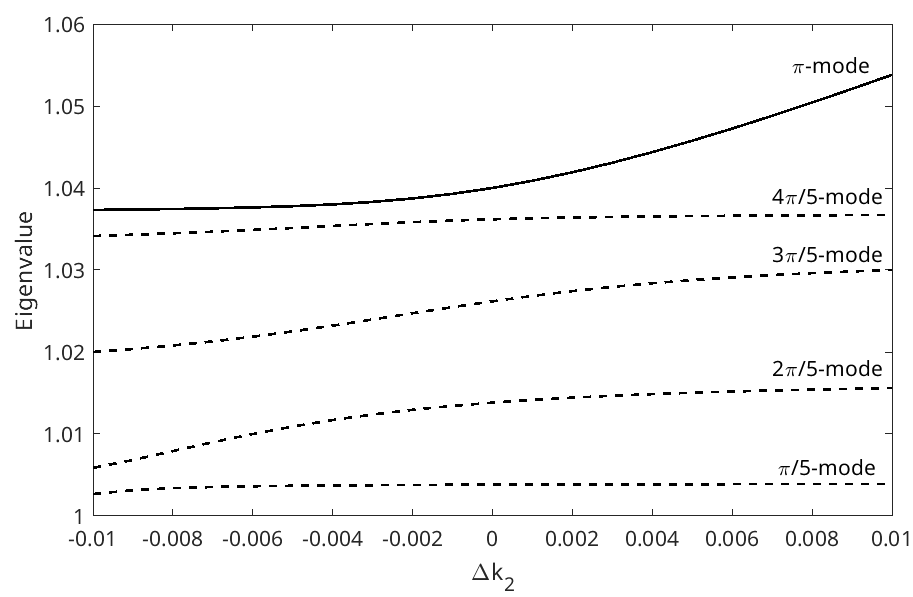}
  \includegraphics[width=0.49\textwidth]{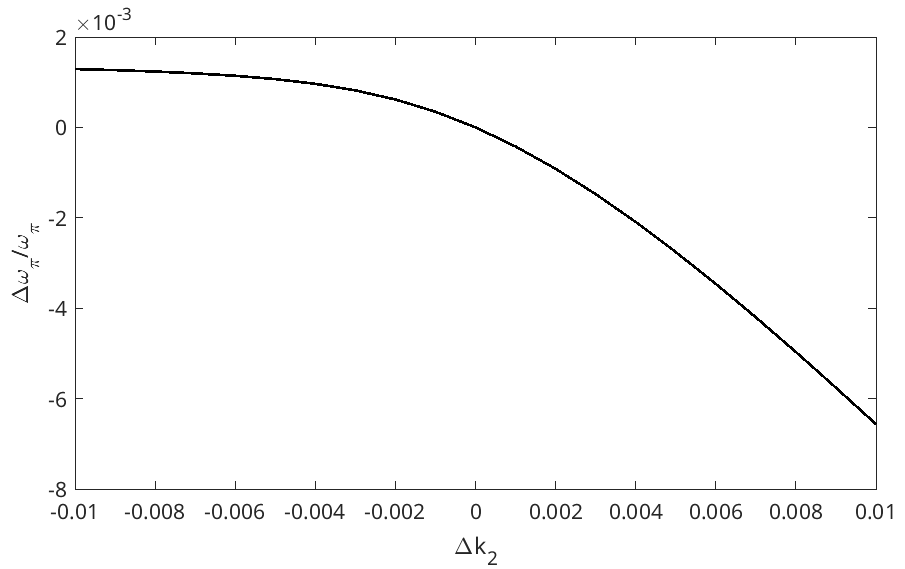}
\end{center}
\caption{\label{fig:dk2}Eigenvalues as function of an error $\Dk_2$ (left)
  and the relative frequency shift $\Delta\omega_{\pi}/\omega_{\pi}$ of the
  $\pi$-mode (right).}
\end{figure}
It is instructive to explore the sensitivity of the eigenfrequencies as a function
of an error of the coupling $\kappa$ in one cell. The right-hand plot in
Figure~\ref{fig:dk2} shows the eigenvalues if the coupling between the second and
third cell $\kappa_2$ varies. The range chosen is rather large; $\Dk_2=-0.01$ corresponds
to absent coupling and $\Dk_2=0.01$ corresponds to a doubling of the design value.
We observe that especially the eigenvalue of the $\pi$-mode increases substantially,
especially for positive values of $\Dk_2$. The right hand-hand plot shows the relative
frequency change $\Delta\omega_{\pi}/\omega_{\pi}$ that is derived from the eigenvalue.
It varies by almost one percent in the range chosen. The slope of the curve around
$\Dk_2=0$ gives us the sensitivity eigenfrequency with respect to $\Dk_2$. From
a fit to scanning $\Dk_2$ over a smaller range, we find
\begin{equation}
  \Delta\left(\Delta\omega_{\pi}/\omega_{\pi}\right)=-0.38\,\Dk_2\ ,
\end{equation}
which is large, considering the very small bandwidth of superconducting cavities.
An error of $\Dk_2=10^{-4}$ thus moves the frequency outside the bandwidth of
the cavity.
\par
The analysis in this section is purely static. In the next sections we will,
however, extend it to account for transient phenomena. In that context we also
introduce specific imperfections to individual cells, which will allow us to
explore their consequences.
\section{Model}
\label{sec:mod}
Let us now consider the current $I_j$ that flows through the $RLC$-branch
of cell~j in the middle of the cavity. The first and last cells need special
considerations and we will come back to them later. The current balance at
the point labeled $V_j$ is given by the sum of the currents passing through
$R$, $L$ and $C$ individually. We thus find~\cite{VZAP}
\begin{equation}\label{eq:RLC}
  0=I_j-I_{j+1}+\frac{V_j}{R_j}+\frac{1}{L_j}\int V_j dt+ C_j\frac{dV_j}{dt}
\end{equation}
The currents $I_j$ and $I_{j+1}$ pass through the capacitors $\hat C_{j-1}$
and $\hat C_j$, respectively and are determined by the difference of voltages
$V_j$ in the adjacent cells
\begin{equation}
  I_j=\hC_{j-1}\left[\frac{dV_j}{dt}-\frac{dV_{j-1}}{dt}\right]
  \qquad\mathrm{and}\qquad
  I_{j+1}=\hC_{j}\left[\frac{dV_{j+1}}{dt}-\frac{dV_{j}}{dt}\right] \ . 
\end{equation}
Inserting in Equation~\ref{eq:RLC} results in
\begin{equation}
  0=-\hC_{j-1}\frac{dV_{j-1}}{dt} -\hC_{j+1}\frac{dV_{j+1}}{dt} + \frac{V_j}{R_j}
  +\frac{1}{L_j}\int V_j dt+ \left(C_j+\hC_{j-1}+\hC_{j+1}\right)\frac{dV_j}{dt}\ .
\end{equation}
Dividing by $C_j$ and introducing the abbreviations $\kappa'_j=\hC_{j-1}/C_j$ and
$\kappa_j=\hC_{j+1}/C_j$ turns this equation into
\begin{eqnarray}\label{eq:cellj}
  0&=&-\kappa'_j\frac{dV_{j-1}}{dt}-\kappa_j\frac{dV_{j+1}}{dt}+ \frac{V_j}{R_jC_j}
       +\frac{1}{L_jC_j}\int V_j dt+ \left(1+\kappa'_j+\kappa_j\right)\frac{dV_j}{dt}
  \nonumber\\
   &=&-\kappa'_j\frac{dV_{j-1}}{dt}-\kappa_j\frac{dV_{j+1}}{dt}+ \frac{\ho_j}{Q_j}V_j
       +\ho_j^2\int V_j dt+ \left(1+\kappa'_j+\kappa_j\right)\frac{dV_j}{dt}\ ,
\end{eqnarray}
where we use the abbreviations $1/L_jC_j=\ho_j^2$ and $R_jC_j=Q_j/\ho_j$ in the
second equality. At this point we introduce the slowly varying amplitude $\tV_j$ of
$V_j=\tV_je^{i\omega t}$, which allows us to write~\cite{VZAP}
\begin{equation}\label{eq:harm}
  \frac{dV_j}{dt}=e^{i\omega t}\left[i\omega\tV_j + \frac{d\tV_j}{dt}\right]
  \qquad\mathrm{and}\qquad
  \int V_j dt = \frac{e^{i\omega t}}{\omega^2}\left[-i\omega\tV_j + \frac{d\tV_j}{dt}\right]\ ,
\end{equation}
for $j=1,\dots,n$. Here we neglect the second derivative $d^2\tV_j/dt^2$ and $\omega$
is the frequency at which the generator operates. After substituting in Equation~\ref{eq:cellj},
canceling the common exponential $e^{i\omega t}$, and some rearranging, we obtain
\begin{eqnarray}\label{eq:tmp1}
  0&=&-\kappa'_j \frac{d\tV_{j-1}}{dt}-\kappa_j \frac{d\tV_{j+1}}{dt}
  +\left(1+\kappa'_j+\kappa_j+\frac{\ho_j^2}{\omega^2}\right)\frac{d\tV_j}{dt} \\
   &&\quad + \frac{\ho_j}{Q_j}\tV_j+i\omega\left[-\kappa'_j\tV_{j-1}-\kappa_j\tV_{j+1} 
      +\left(1+\kappa'_j+\kappa_j-\frac{\ho_j^2}{\omega^2}\right)\tV_j\right]\ .\nonumber
\end{eqnarray}
The quantity in the square brackets turns out to be the equivalent of Equation~\ref{eq:cj}
provided the all $\kappa'_j$ and $\kappa_j$ are equal. The value of $\omega$ for which this
bracket vanishes thus determines the eigenvalues of the matrix in Equation~\ref{eq:evs}
that describe the steady-state of the system, because in that case, also the derivatives
of the voltages with respect to time have to vanish. The additional term, proportional to
$\ho_j/Q_j$, is due to the finite resistance $R_j$ and describes damping. This term is absent
in the discussion in Section~\ref{sec:ss}. For future reference, we rewrite Equation~\ref{eq:tmp1}
in the following way
\begin{eqnarray}\label{eq:cjt}
  &&-\kappa'_j \frac{d\tV_{j-1}}{dt}
     +\left(1+\kappa'_j+\kappa_j+\frac{\ho_j^2}{\omega^2}\right)\frac{d\tV_j}{dt}
  -\kappa_j \frac{d\tV_{j+1}}{dt}\\
   &&\quad = -\frac{\ho_j}{Q_j}\tV_j-i\omega\left[-\kappa'_j\tV_{j-1}
      +\left(1+\kappa'_j+\kappa_j-\frac{\ho_j^2}{\omega^2}\right)\tV_j-\kappa_j\tV_{j+1} \right]\ ,\nonumber
\end{eqnarray}
which separates the terms with the derivatives on the left-hand side of the equality
and the rest on the right-hand side.
\par
For the last cell~$n$ we start from the current balance on the node labeled $V_n$ in
Figure~\ref{fig:mcc}
\begin{equation}
  0=I_n-I_{n+1}+\frac{V_n}{R_n} +\frac{1}{L_n}\int V_ndt + C_n\frac{dV_n}{dt}
\end{equation}
and express the currents through the voltages in the adjacent nodes
\begin{equation}
  I_n=\hC_{n-1}\left[\frac{dV_j}{dt}-\frac{dV_{j-1}}{dt}\right]
  \qquad\mathrm{and}\qquad
  I_{n+1}=C_b\left[0-\frac{dV_{n}}{dt}\right] \ , 
\end{equation}
where the capacitor $C_b$ is connected to ground potential on its other end, which
accounts for the zero in the square bracket. Collecting terms and introducing the
abbreviations $\kappa'_n=\hC_{n-1}/C_n$ and $\kappa_b=C_b/C_n$ leads to
\begin{equation}
  0=-\kappa'_n \frac{dV_{n-1}}{dt}+\left(1+\kappa'_n+\kappa_b\right)  \frac{dV_{n}}{dt} 
  + \frac{\ho_n}{Q_n}V_n +\ho_n^2\int V_n dt\ , 
\end{equation}
where we introduced $1/L_nC_n=\ho_n^2$ and $R_nC_n=Q_n/\ho_n$. With the help of
Equation~\ref{eq:harm}, and after canceling the common exponential $e^{i\omega t}$,
this equation becomes
\begin{eqnarray}
  0&=&-\kappa'_n \frac{d\tV_{n-1}}{dt}
  +\left(1+\kappa'_n+\kappa_b+\frac{\ho_n^2}{\omega^2}\right)\frac{d\tV_j}{dt} \\
   &&\quad + \frac{\ho_n}{Q_n}\tV_n+i\omega\left[-\kappa'_n\tV_{n-1}
      +\left(1+\kappa'_n+\kappa_b-\frac{\ho_n^2}{\omega^2}\right)\tV_n\right]\ .\nonumber
\end{eqnarray}
As before, the quantity in the square bracket corresponds to Equation~\ref{eq:cn} and
gives rise to the bottom row in the matrix from Equation~\ref{eq:evs}. Some
rearranging then leads to
\begin{eqnarray}\label{eq:cnt}
  &&-\kappa'_n \frac{d\tV_{n-1}}{dt}
  +\left(1+\kappa'_n+\kappa_b+\frac{\ho_n^2}{\omega^2}\right)\frac{d\tV_j}{dt} \\
   &&\qquad = -\frac{\ho_n}{Q_n}\tV_n -i\omega\left[-\kappa'_n\tV_{n-1}
      +\left(1+\kappa'_n+\kappa_b-\frac{\ho_n^2}{\omega^2}\right)\tV_n\right]\ ,\nonumber
\end{eqnarray}
where we also collect the derivatives on the left-hand side.
\par
Finally, we come to the first cell, which is also connected to the coupler on top
of the capacitances $\hC_1$ and $C_a$. The current balance at the node labeled $V_1$
for this cell is
\begin{eqnarray}
  \frac{I_g}{n} &=& \frac{V_1}{n^2Z_0} + \frac{V_1}{R_1}+\frac{1}{L_1}\int V_1dt
       + \left(C_1+C_a\right)\frac{dV_1}{dt} -I_2\\
   &=& \frac{I_g}{n}+\left( 1+\beta\right) \frac{V_1}{R_1} +\frac{1}{L_1}\int V_1dt
       + \left(C_1+C_a\right)\frac{dV_1}{dt} -\hC_1\left[\frac{dV_2}{dt}-\frac{dV_1}{dt}\right]\ ,
       \nonumber
\end{eqnarray}
where we introduce the coupling factor $\beta=R_1/n^2Z_0$. Note that we have to transform
the impedance $Z_0$ through the transformer to the inside of the cavity, which accounts
for the factor $n^2$ in the denominator and dividing the generator current $I_g$ by $n$
on the ``other'' side of the transformer. Furthermore, dividing this equation by $C_1$
and introducing the abbreviations $\kappa_1=\hC_1/C_1$ and $\kappa_a=C_a/C_1$,
we arrive at
\begin{equation}
 \frac{1}{C_1}\frac{I_g}{n} =(1+\beta)\frac{\ho_1}{Q_1}V_1
  +\ho_1^2\int V_1dt+\left(1+\kappa_a+\kappa_1\right)\frac{dV_1}{dt} -\kappa_1\frac{dV_2}{dt}\ .
\end{equation}
After using Equation~\ref{eq:harm} and canceling the common exponential $e^{i\omega t}$,
we obtain
\begin{eqnarray}\label{eq:c1ttemp}
  \frac{1}{C_1}\frac{\tilde I_g}{n}&=&\left(1+\kappa_a+\kappa_1+\frac{\ho_1^2}{\omega^2}\right)\frac{d\tV_1}{dt}
       -\kappa_1\frac{d\tV_2}{dt}\\
     &&\qquad  +(1+\beta)\frac{\ho_1}{Q_1}\tV_1
        +i\omega\left[\left(1+\kappa_a+\kappa_1-\frac{\ho_1^2}{\omega^2}\right)\tV_1
        -\kappa_1\tV_2\right]\ .\nonumber
\end{eqnarray}
The square bracket describes Equation~\ref{eq:c1} and corresponds to the first row of the
matrix in Equation~\ref{eq:evs}. Now we introduce the loaded $Q$-value $Q'_L=Q_1/(1+\beta)$
and note that $1/C_1=\ho_1R_1/Q_1=\ho_1R_1/(1+\beta)Q'_L$. Here $Q'_L$ is the loaded $Q$-value
for the first cell only. In passing, we point out that instead of $Q_1$, the expression
$(1+\beta)Q'_L$ often appears in the literature. After rearranging Equation~\ref{eq:c1ttemp},
we arrive at
\begin{eqnarray}\label{eq:c1t}
  &&\left(1+\kappa_a+\kappa_1+\frac{\ho_1^2}{\omega^2}\right)\frac{d\tV_1}{dt}
  -\kappa_1\frac{d\tV_2}{dt}\\
     &&\qquad=-\frac{\ho_1}{Q'_L}\tV_1
        -i\omega\left[\left(1+\kappa_a+\kappa_1-\frac{\ho_1^2}{\omega^2}\right)\tV_1
        -\kappa_1\tV_2\right] + \frac{\ho_1R_1}{Q_1}\frac{\tilde I_g}{n}\ .\nonumber
\end{eqnarray}
In the absence of coupling we only have a single cell and a single mode with $\omega\approx\ho_1$
to deal with. In that case, this equation reverts to $2d\tV_1/dt=-(\ho_1/Q'_L)\tV_1-i\ho_1\delta_1\tV_1
+\ho_1R_1/((1+\beta)Q'_L){\tilde I_g}/n$ with the detuning $\delta_1=\omega/\ho_1-\ho_1/\omega$,
which agrees with the calculation for a single resonator in~\cite{VZAP}.
\par
Equations~\ref{eq:cjt}, \ref{eq:cnt}, and \ref{eq:c1t} describe the dynamics for multi-cell
cavity when excited by the generator with frequency~$\omega$. In the following sections we
will consider the most important case, when the cavity is driven such that the $\pi$-mode
is excited.
\section{Perfect $\pi$-mode cavity}
Here we assume that the frequency at which the generator operates is $\opi=\ho/\sqrt{1+4\kappa}$,
where $\ho$ is the design value of a single cell's resonance frequency and $\kappa$ is the
design value of the coupling between adjacent cells. $Q$ is the unloaded quality factor of
the cells, all assumed to be equal. With the design values for $\kappa_a=\kappa_b=2\kappa$ we
can transform Equation~\ref{eq:c1t}, describing the first cell, into
\begin{eqnarray}\label{eq:c1tp}
\left(2+7\kappa\right)\frac{d\tV_1}{dt} -\kappa\frac{d\tV_2}{dt}
     &=&-\frac{\ho}{Q'_L}\tV_1
        -i\opi\left[-\kappa\tV_1
         -\kappa\tV_2\right] + \frac{\ho R_1}{Q_1}\frac{\tilde I_g}{n}\\
     &=&-\frac{\ho}{Q'_L}\tV_1
         +i\frac{\ho\kappa}{\sqrt{1+4\kappa}}\left[\tV_1+\tV_2\right]
         + \frac{\ho R_1}{Q_1}\frac{\tilde I_g}{n}\ .\nonumber
\end{eqnarray}
with $Q'_L=Q/(1+\beta)$. Likewise, the equation for cell~$j$ with $2\leq j \leq n-1$ in the
middle of the cavity is based on Equation~\ref{eq:cjt}, which transforms into
\begin{eqnarray}\label{eq:cjtp}
&&  -\kappa \frac{d\tV_{j-1}}{dt} +\left(2+6\kappa\right)\frac{d\tV_j}{dt}
     -\kappa \frac{d\tV_{j+1}}{dt}\nonumber\\
   &&\qquad\qquad= -\frac{\ho}{Q}\tV_j-i\opi\left[-\kappa\tV_{j-1}
      -2\kappa\tV_j-\kappa\tV_{j+1} \right] \\
  &&\qquad\qquad = -\frac{\ho}{Q}\tV_j+i\frac{\ho\kappa}{\sqrt{1+4\kappa}}
     \left[\tV_{j-1} +2\tV_j+\tV_{j+1} \right]\ .\nonumber
\end{eqnarray}
Finally, for the last cell~$n$, we utilize Equation~\ref{eq:cnt}, which transforms into
\begin{eqnarray}\label{eq:cntp}
  -\kappa \frac{d\tV_{n-1}}{dt}+\left(2+7\kappa\right)\frac{d\tV_j}{dt} 
  &=& -\frac{\ho}{Q}\tV_n -i\opi\left[-\kappa\tV_{n-1}-\kappa\tV_n\right]\\
  &=& -\frac{\ho}{Q}\tV_n +i\frac{\ho\kappa}{\sqrt{1+4\kappa}}
      \left[\tV_{n-1}+\tV_n\right]\nonumber
\end{eqnarray}
These three equations thus describe the transient behavior of a perfect multi-cell
cavity operating in the $\pi$-mode.
\par
For the simulations, we multiply the equation by $\dt$, which makes it numerically
more stable, because then most factors are of magnitude unity. Moreover, we
transform Equations~\ref{eq:c1tp}, \ref{eq:cjtp}, and \ref{eq:cntp} into the
following matrix-valued equation
\begin{equation}\label{eq:dynp}
  D\frac{d\vec V}{dt}\dt = -(A-iB)\vec V +\frac{\ho\dt R_1}{Q_1}\vec I
\end{equation}
where we used the abbreviations
\begin{eqnarray}\label{eq:dynpaux}
  \vec V &=& \left(\tV_1,\tV_2,\tV_3,\tV_4,\tV_5\right)^{\top}\nonumber\\
  \frac{d\vec V}{dt} &=&
       \left(\frac{d\tV_1}{dt} , \frac{d\tV_2}{dt} , \frac{d\tV_3}{dt} , 
       \frac{d\tV_4}{dt} , \frac{d\tV_5}{dt} \right)^{\top}\nonumber\\
  \vec I &=& \left(\tilde I_g/n,0,0,0,0\right)^{\top}\nonumber\\
  D&=&\left(\begin{array}{ccccc}
     2+7\kappa & -\kappa & 0 & 0 & 0 \\
     -\kappa & 2+6\kappa & - \kappa & 0 & 0 \\
     0 & -\kappa & 2+6\kappa & - \kappa & 0 \\
     0 & 0 & -\kappa & 2+6\kappa & - \kappa \\
     0 & 0 &  0 &-\kappa & 2+7\kappa
   \end{array}\right)\\
  A&=&\diag\left((1+\beta)\frac{\ho\dt}{Q}, \frac{\ho\dt}{Q}, \frac{\ho\dt}{Q},
       \frac{\ho\dt}{Q}, \frac{\ho\dt}{Q}\right)\nonumber\\
  B&=&\frac{\ho\dt\kappa}{\sqrt{1+4\kappa}}\left(\begin{array}{ccccc}
              1 & 1 & 0 & 0 & 0\\
              1 & 2 & 1 & 0 & 0\\
              0 & 1 & 2 & 1 & 0\\
              0 & 0 & 1 & 2 & 1\\
              0 & 0 & 0 & 1 & 1
 \end{array}\right)\ . \nonumber
\end{eqnarray}
We further manipulate the system by introducing real and imaginary parts of voltages 
and currents
\begin{equation}
  \vec V = \vec V_r + i\vec V_i
  \qquad\mathrm{and}\qquad
  \vec I = \vec I_r + i\vec I_i \ . 
\end{equation}
This transforms Equation~\ref{eq:dynp} into
\begin{equation}\label{eq:dynpb}
  \left(\begin{array}{cc} D & 0_5 \\ 0_5 & D\end{array}\right)
  \left(\begin{array}{c} \frac{d\vec V_r}{dt} \\ \frac{d\vec V_i}{dt}\end{array}\right)\dt
  =-\left(\begin{array}{rr} A & B \\ -B & A\end{array}\right)
  \left(\begin{array}{c} \vec V_r \\ \vec V_i\end{array}\right)
  +\frac{\ho\dt R_1}{Q_1} \left(\begin{array}{c} \vec I_r \\ \vec I_i\end{array}\right)\ ,
\end{equation}
where $0_5$ is a $5\times 5$ matrix containing only zeros. Here we are now dealing
with the 10 real and imaginary parts, or equivalently the I and Q components, of
the voltages in the five cells. 
\par
The steady-state conditions are given by vanishing time derivatives on the left-hand
side and that leads to the steady-state voltages
\begin{equation}
  \tilde V_{\infty} =
  \left(\begin{array}{c} \vec V_r \\ \vec V_i\end{array}\right)_{\infty}
  = \frac{\ho\dt R_1}{Q_1}
  \left(\begin{array}{rr} A & B \\ -B & A\end{array}\right)^{-1}
  \left(\begin{array}{c} \vec I_r \\ \vec I_i\end{array}\right)\ .
\end{equation}
where we introduce $\tilde V=(\vec V_r,\vec V_i)^{\top}$.
\par
We obtain the temporal evolution of the voltages $\tilde V$ as a function of the
re-scaled time $s=t/\dt$ from Equation~\ref{eq:dynpb}
\begin{equation}\label{eq:dynpc}
  \frac{d\tilde V}{ds} = -\tilde A \tilde V(s)+\frac{\ho\dt R_1}{Q_1}\tilde I
  \quad\mathrm{with}\quad
  \tilde A=\left(\begin{array}{rr} D^{-1}A & D^{-1}B \\ -D^{-1}B & D^{-1}A\end{array}\right)
\end{equation}
and $\tilde I(s)=(D^{-1}I_r, D^{-1}I_i)^{\top}$. Note that here $\tilde V$ and $\tilde I$
depend on $s$.
\par
In our case $\tilde A$ and $\tilde I$ are constant, so we can formally
integrate Equation~\ref{eq:dynpc} with the result
\begin{equation}
  \tilde V(s)=\exp\left[-\tilde A s\right]\tilde K + \tilde V_{\infty}\ ,
\end{equation}
where $\tilde K$ is a vector with the integration constants for the voltages.
Matching the initial conditions $\tilde V(s=0)=\tilde V_0$ leads us to
$\tilde K=\tilde V_0-\tilde V_{\infty}$ and that gives us the formal solution 
\begin{equation}\label{eq:dynpd}
  \tilde V(s)=\exp\left[-\tilde A s\right]\left(\tilde V_0-\tilde V_{\infty}\right)
  + \tilde V_{\infty}\ ,
\end{equation}
where the different $\tilde V$ are ten-component vectors with the five
real parts as upper entries and the five imaginary parts just below. The
matrix-exponent $\exp\left[-\tilde A s\right]$ is a $10\times 10$ matrix
that can be calculated from the eigenvalue decomposition of~$\tilde A$.
\par
Let us assume that we write $\tilde A$ expressed through its eigenvalues
and eigenvectors, assembled in $W$, as
\begin{equation}\label{eq:WLW}
  \tilde A= W\Lambda W^{-1}
  \qquad\mathrm{with}\qquad
  \Lambda=\diag\left(\lambda_1,\dots,\lambda_{10}\right)\ .
\end{equation}
Now let us imagine $\exp\left[-\tilde A s\right]$ being expressed as a power series
of $\tilde A s$. Then it is easy to see that calculating the powers of $\tilde A s$
always brings $W$ and $W^{-1}$ next to each other, so they cancel, leaving only
the first and last $W$ and $W^{-1}$ with powers of $(\Lambda s)$ in-between.
And they combine to the form the matrix exponential
\begin{equation}\label{eq:expm}
  \exp\left[-\tilde A s\right]=W\left(e^{-\lambda_1s},\dots,e^{-\lambda_{10}s}\right) W^{-1}\ ,
\end{equation}
which provides us with a recipe to calculate an explicit representation of
the matrix exponential and to numerically evaluate Equation~\ref{eq:dynpd}.
All we need is the eigenvalue decomposition of $\tilde A$ and that is easily
done with Matlab's built-in function {\tt eig()}. This way of calculating the
voltages is very efficient; we only need to call {\tt eig()} once and then, for
each time $s$, first evaluate Equation~\ref{eq:expm} and then Equation~\ref{eq:dynpd}.
We implement the algorithm in a Matlab script and run it for 10000 time steps of
duration $\dt$. In the simulations we use coupling $\beta=5000$, unloaded
$Q=10^9$, shunt impedance $R=10^6\,\Omega$, resonance frequency $\ho/2\pi=1.5\,$GHz,
and cell-to-cell coupling $\kappa=0.01$. Moreover, we use $I_g/n=1\,$kA.
\par
\begin{figure}[tb]
\begin{center}
\includegraphics[width=0.49\textwidth]{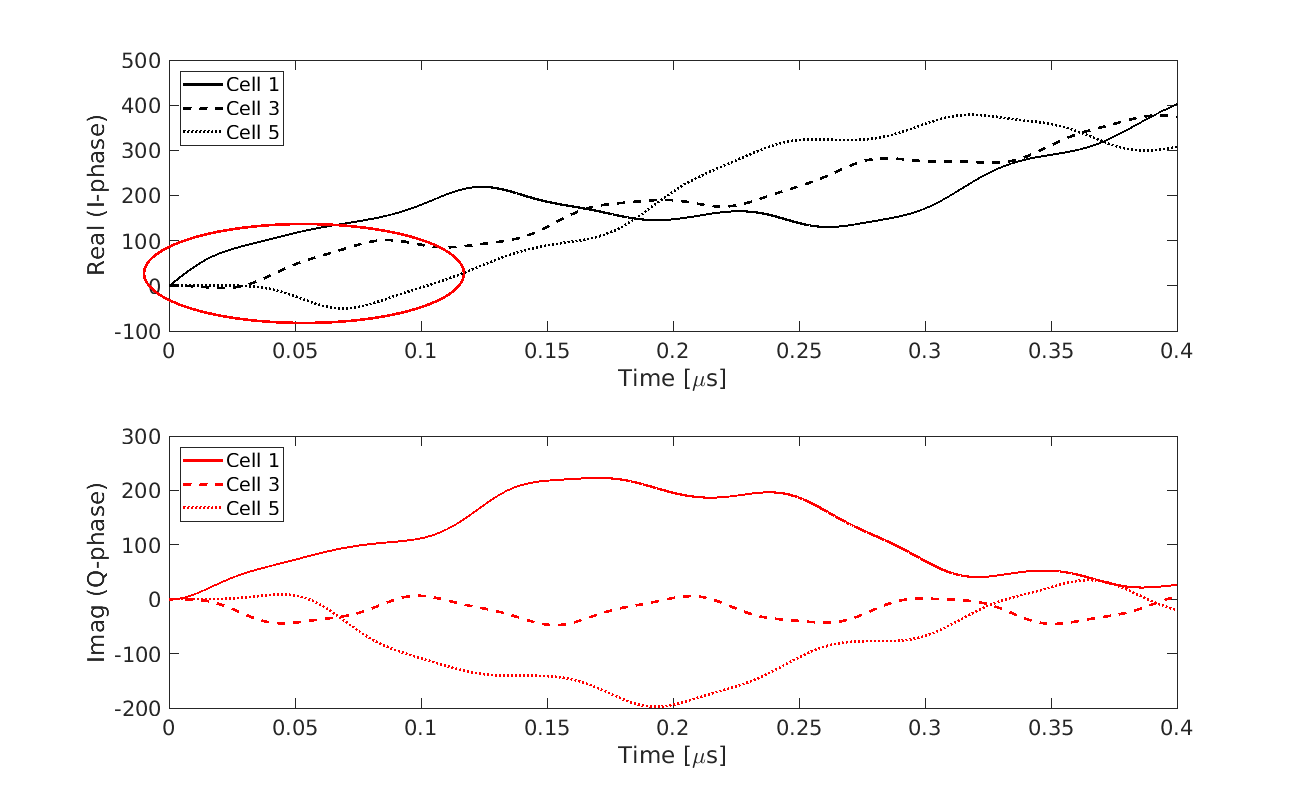}
\includegraphics[width=0.49\textwidth]{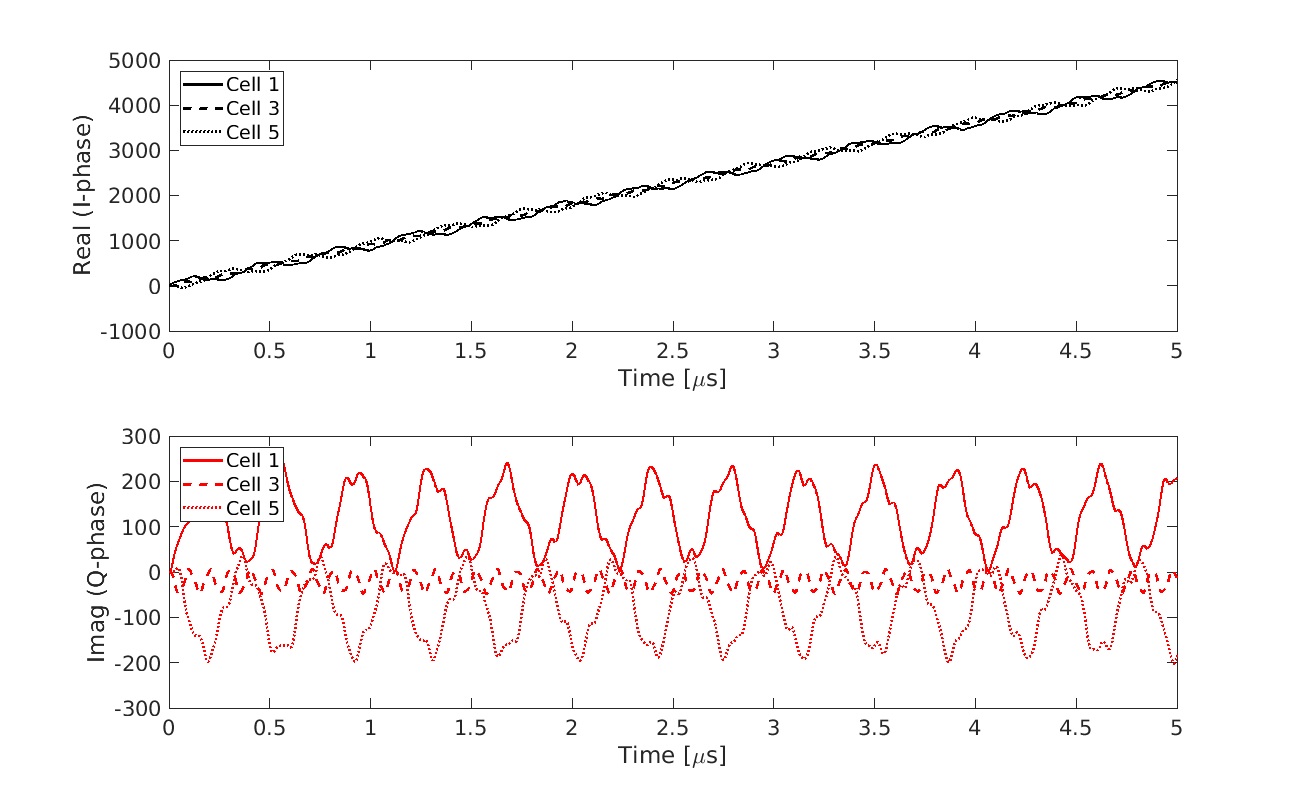}
\end{center}
\caption{\label{fig:fivea}Left: real (top) and imaginary (bottom) phase in
  cells~1, 3, and 5 for the initial 0.4\,\textmu s after the generator is switched on.
  Right: the corresponding signals for the first 5\,\textmu s.}
\end{figure}
The left-hand panel in Figure~\ref{fig:fivea} shows the real part
of the voltages in cells~1, 3, and 5 in the upper plot and the corresponding
imaginary parts in the lower plot. We omitted the fields in cells~2 and~4,
because they have opposite polarity to the odd-numbered cells in the $\pi$-mode
and showing them would make the plot
overly complex. In this simulation we use $\dt=4\times 10^{-11}\,$s such that
the horizontal time axis extends to 0.4\,\textmu s. In the upper plot, we used
the red ellipse to highlight the startup process with the field in cell~1 (solid)
increases first, followed by the field in cell~3 (dashed) and a little later also
the field in cell~5 (dotted) increases. But already after 0.2\,\textmu s, the field
starts to exhibit an overall increase. The imaginary phase, shown in the bottom plot,
only shows beating back and forth, but no overall increase.
\par
The right-hand panel shows a simulation where the time step is increased to
$\dt=5\times10^{-10}\,$s such that the horizontal axis extends to 5\,\textmu s.
Now the real part of the fields in the three cells, shown in the upper plot,
clearly exhibits a linear increase. Only a very small agitation around the linear
rise is visible. The imaginary parts of the voltages, shown on the bottom plot,
show some beating, though on a much reduced scale compared to
the upper plot. Already here we can conclude that the fields in the odd-numbered
cells rises in unison which indicates that the generator actually drives the
$\pi$-mode as an entity, rather than first exciting cell~1 that in turn excites
cell~2 and so forth. The equilibration among cells happens on a time scale that
is faster than one microsecond.
\par
One might wonder, whether the oscillations seen in Figure~\ref{fig:fivea} can be
directly observed, at least on the transmitted and maybe on the reflected signal.
They happen on a time scale of 100\,ns and measuring them 
would require a data-acquisition bandwidth of 10\,MHz or more. Normal signal processing
typically uses a much smaller bandwidth such that these oscillations are averaged
out and are likely not observed. On the other hand, exploring them further for
diagnostic purposes might prove interesting and we will pursue this further in
the future.
\par
\begin{figure}[tb]
\begin{center}
\includegraphics[width=0.9\textwidth]{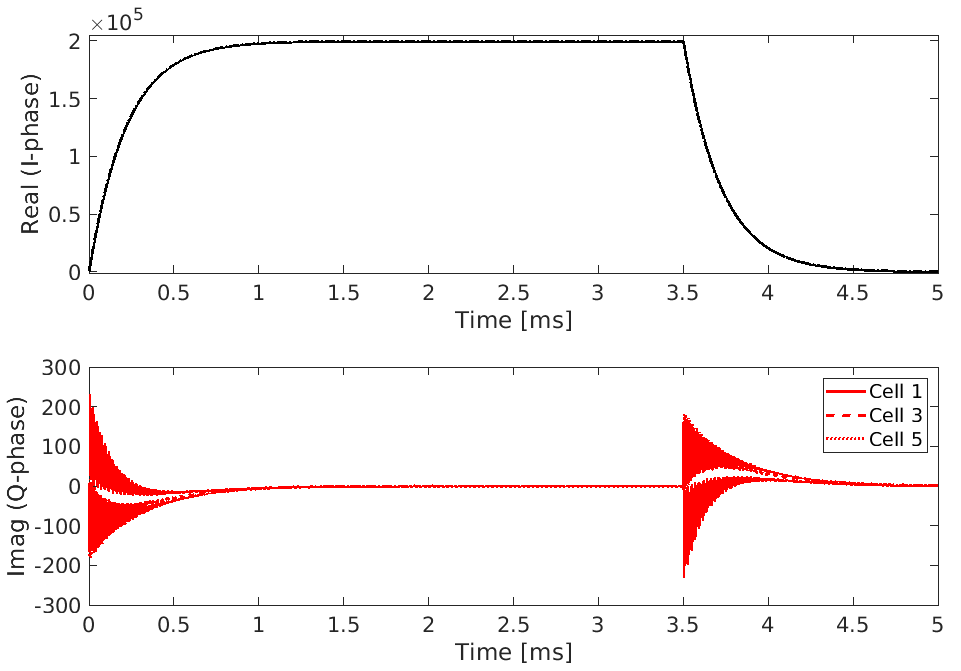}
\end{center}
\caption{\label{fig:fiveb}Real (top) and imaginary (bottom) phase in
  cells~1, 3, and 5. The generator is switched on between zero and 3.5\,ms.}
\end{figure}
\begin{figure}[tb]
\begin{center}
\includegraphics[width=0.9\textwidth]{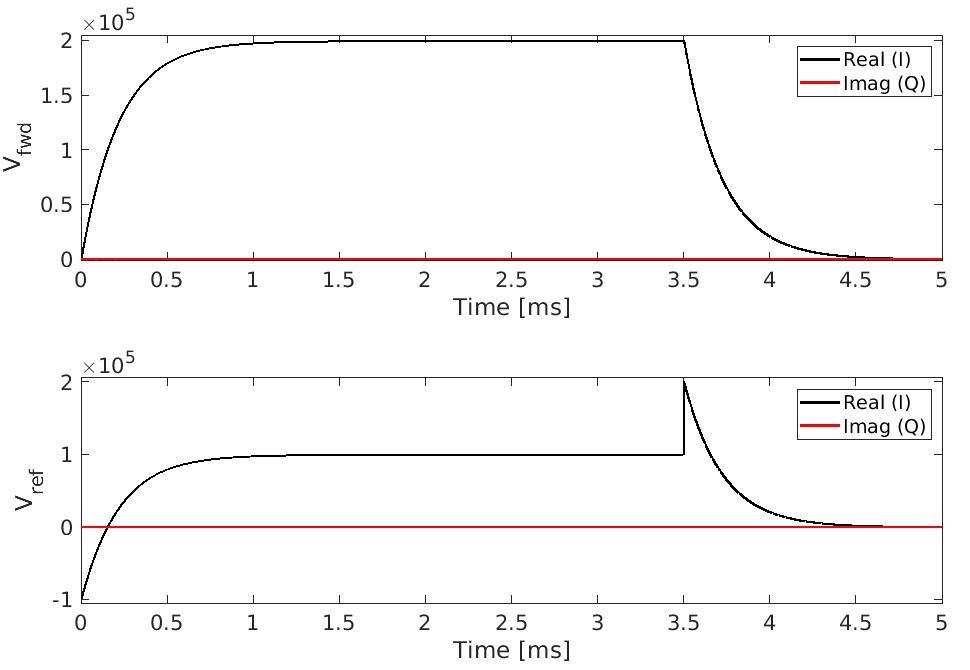}
\end{center}
\caption{\label{fig:fivec}Top: real and imaginary phase of the transmitted signal,
  assumed to be picked up in cell~5. Bottom: real and imaginary phase of the
  reflected signal. The generator is switched on between zero and 3.5\,ms.}
\end{figure}
Figure~\ref{fig:fiveb} shows the fields coming from a simulation with
$\dt=5\times 10^{-7}\,$s such that the horizontal axis extends to 5\,ms.
Apart from turning on the generator at time zero, we again turn it off
after 3.5\,ms. As before, the real parts of the fields are shown in the
upper plot. On this time scale, differences between the voltages in
cells~1, 3, and~5 are not discernible. 
They follow the expected behavior of first filling the
cavity and, after the generator is turned off after 3.5\,ms, emptying
it. We observe quite some agitation on the lower plot immediately after
turning the generator on at time zero and likewise after turning it off
after 3.5\,ms, albeit at a much lower amplitude compared to the upper plot.
\par
The upper plot in Figure~\ref{fig:fivec} shows the real and imaginary part of
the voltage in the fifth cell $V_5$, which is assumed to be equipped with a field
probe that would provide the transmitted signal. The lower plot shows the
real and imaginary part of the reflected voltage pulse that travels back
towards the generator. It is given by~\cite{VZAP}
\begin{equation}
  nV_{ref}=V_1-\frac{R_1I_g/n}{2\beta}\ ,
\end{equation}
where $V_1$ is the voltage in cell~1. This signal exhibits the distinct negative
spike at the start of the pulse and a positive spike at its end.
\par
\begin{figure}[p]
\begin{center}
  \includegraphics[width=0.65\textwidth]{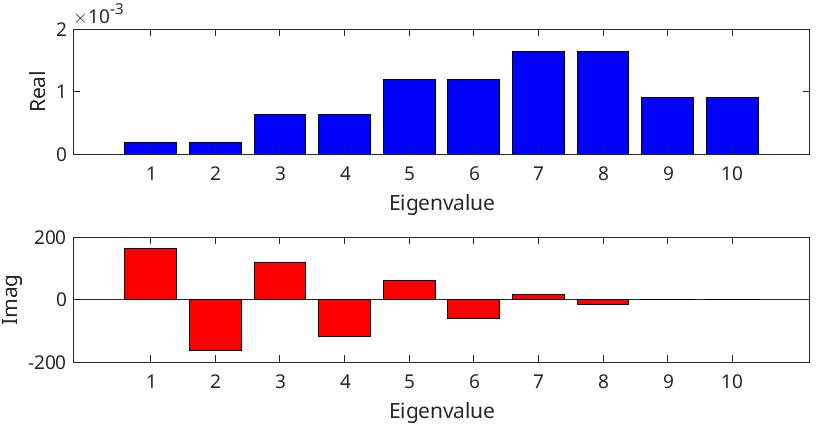}
  \caption{\label{fig:eigval}Real(top) and imaginary (bottom) parts
    of the eigenvalues of $\tilde A$.}
  \vskip 5mm
  \includegraphics[width=0.65\textwidth]{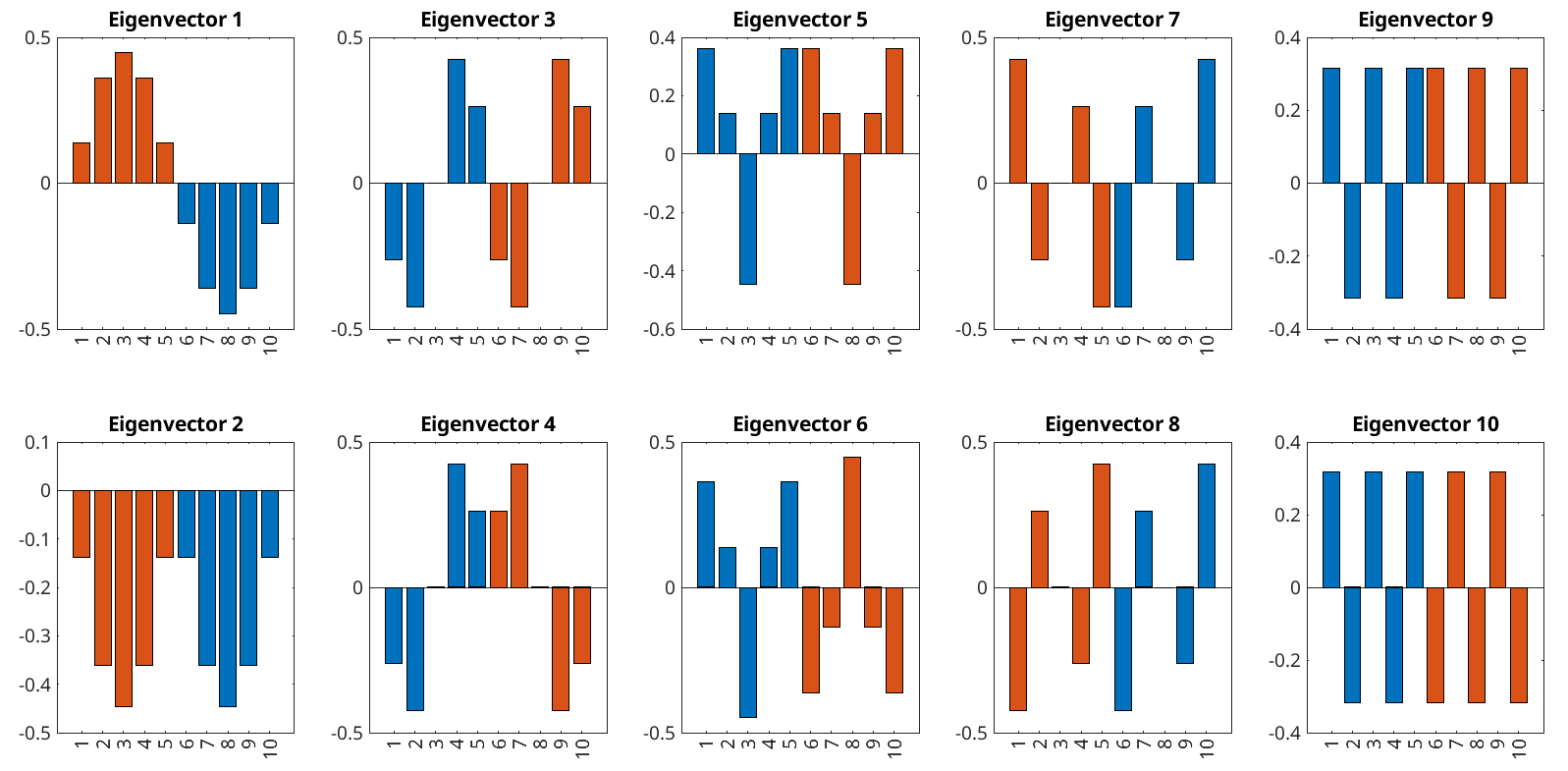}
  \caption{\label{fig:eigvec}Eigenvectors of $\tilde A$.}
  \vskip 5mm
  \includegraphics[width=0.65\textwidth]{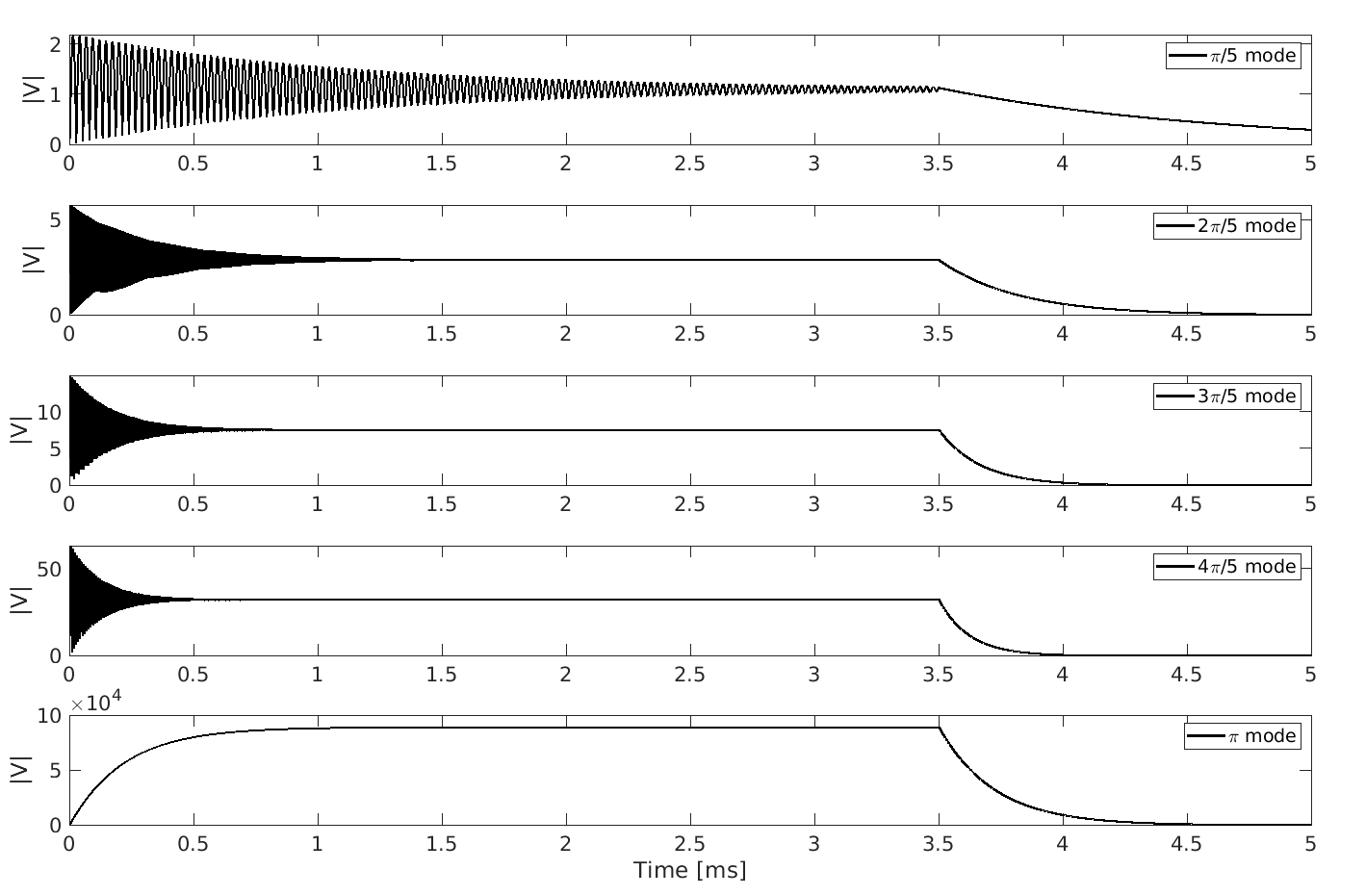}
  \caption{\label{fig:modamp}Mode amplitudes. Note the wide range of
    vertical scales for the five plots. The $\pi$-mode, shown on the
    bottom plot, reaches a much larger amplitude than the other four
    plots.} 
\end{center}
\end{figure}
From the discussion immediately following Equation~\ref{eq:c1t} one might expect
that the filling time $\tau$ of the cavity is given by $1/\tau=\ho_1/2Q'_L$.
In reality, however, it is determined by the real part of the eigenvalue that
corresponds to the $\pi$-mode excitation of the cavity. In order to explore this
further we display the ten eigenvalues in Figure~\ref{fig:eigval}. The upper plot
shows the real parts of the eigenvalues and the bottom plot shows the corresponding
imaginary parts, which is much larger than the corresponding real parts. We also
observe that the eigenvalues come in complex-conjugate pairs; adjacent real parts
have the same magnitude and the corresponding imaginary parts have opposite sign.
Here a large imaginary part indicates that the mode is mostly oscillating. They
are responsible for the oscillatory behavior of the cavity voltages at the start
of the pulse, highlighted in Figure~\ref{fig:fivea}. The real parts of eigenvalues~1
and~2 are particularly small, so they cause oscillations of the corresponding mode
to linger on for a relatively long time. At the same time, their imaginary part is
large, such that they are oscillating substantially. This explains the excitation
of the imaginary part, the Q-phase, in Figure~\ref{fig:fiveb}. The reason that the
first eight eigenvalues are excited at all comes from the sudden rise of the pulse
from the generator, which contains many frequencies. And some of them are inside
the bandwidth of the first four modes that are described by the first eight eigenvalues.
On the other hand, the pair of eigenvalues numbered~9 and~10 show the smallest
imaginary part (red) while still having a moderate real part (blue). These two
modes therefore mostly absorb the energy delivered by the generator.
\par
The upper row in Figure~\ref{fig:eigvec} shows the odd-numbered eigenvectors. The
real parts are shown in blue and the imaginary parts in red. That they come in groups
of five is a consequence of ordering the voltages when transforming the complex-valued
Equation~\ref{eq:dynp} to its real-valued counterpart Equation~\ref{eq:dynpb}. We
observe that the modal pattern follows the modes already seen in Figure~\ref{fig:evs}.
Only here the real (blue) and imaginary (red) parts mimic each other. The
bottom row shows the complex-conjugate even-numbered eigenvectors. They show the
same real part (blue) as the corresponding plot above, but the sign of the imaginary
part (red) is reversed.
\par
In particular, eigenvectors~9 and~10 display the excitation pattern of the $\pi$-mode
with adjacent cells having opposite polarity. Notably, we already found that the
imaginary part of the corresponding eigenvalues~9 and~10 is particularly small and
therefore, rather than oscillating, this mode absorbs most of the energy from the
generator. And the real part of the eigenvalues determines the filling time
$\tau=1/\real(\lambda_9)$ of the
$\pi$-mode. Comparing its numerical value with $1/\tau'= \ho/2Q'_L$ we find that
$\tau=5\tau'$. It takes five times longer to fill a five-cell cavity compared to
filling a single cell. After all, we have to provide power to fill five cells
rather than one, and that takes five times longer. Using a complementary point
of view: the five cells behave like five bandpass filters in series and therefore
the overall bandwidth is reduced five-fold.
\par
In order to see the temporal dependence of the five modes, we create projectors
onto their corresponding eigenspaces. These eigenspaces are spanned by the
columns of $W$, which are the eigenvectors of $W$ from Equation~\ref{eq:WLW}.
Therefore, we can write
\begin{equation}
W=\left(\ket{w_1},\ket{w_2},\dots,\ket{w_{10}}\right)\ .
\end{equation}
Here we borrow the notation with bra and ket vectors from quantum mechanics,
where ket denotes column vectors, the eigenvectors $\ket{w_i}$, $i=1,\dots,10$
that are the columns of the matrix $W$. In a similar fashion, we denote the
row vectors of $W^{-1}$ by the $\bra{w_i}$, thus
\begin{equation}
W^{-1}=\left(\begin{array}{c}\bra{w_1}\\ \vdots \\ \bra{w_{10}}\end{array}\right)\ .
\end{equation}
Using this notation, we can rewrite Equation~\ref{eq:WLW} as a spectral
decomposition
\begin{equation}
  \tilde A =\sum_{i=1}^{10}\lambda_i P_i
  \qquad\mathrm{with}\qquad
  P_i=\ket{w_i}\bra{w_i}\ ,
\end{equation}
where $P_i$ is the projector onto the subspace characterized by eigenvalue
$\lambda_i$ and $\ket{w_i}\bra{w_i}$ is the outer product of column vector
$\ket{w_i}$ and the row vector $\bra{w_i}$. Note that $\tilde A$ is not
symmetric and therefore $W$ is neither real nor orthogonal. Since the eigenvalues
come in complex conjugate pairs, we use the sum of $P_1$ and $P_2$ or $Q_1=P_1+P_2$,
for example, to determine the contribution of the $\pi/5$-mode to the voltage
$\tilde V(s)$, which is given by $V(\pi/5)=Q_1\tilde V(s)$. Building the
projectors for the other modes works analogously. The upper plot
in Figure~\ref{fig:modamp} shows the amplitude of this mode, given by the
norm of $V(\pi/5)$. We observe that it only slowly decays owing to the fact that
eigenvalues $\lambda_1$ and $\lambda_2$ have the smallest real part and therefore
correspond to the longest time constant, as already discussed above. All modes,
except the $\pi$-mode shown on the bottom plot, decay to small values long
before the generator is turned off at 3.5\,ms. Only the amplitude of the
$\pi$-mode increases to much larger values than the other four modes and
follows the behavior already seen on the top plots in Figures~\ref{fig:fiveb}
and~\ref{fig:fivec}.
\par
\begin{figure}[b]
\begin{center}
  \includegraphics[width=0.8\textwidth]{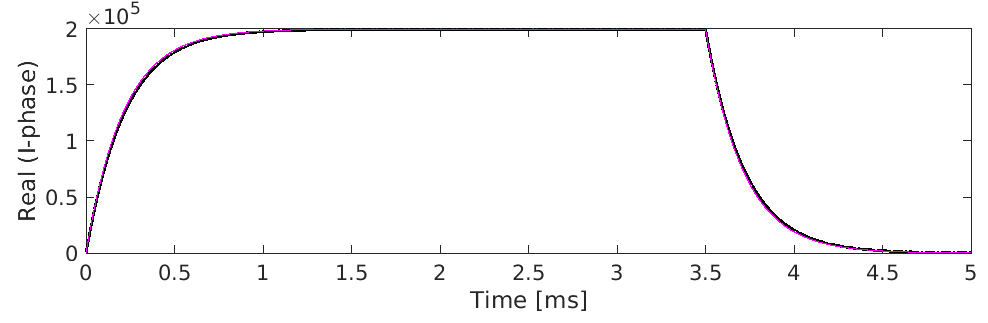}
  \caption{\label{fig:comp}The voltage in the single cell cavity (magenta) superimposed
    to the voltage in the five-cell cavity (black).} 
\end{center}
\end{figure}
It is instructive to compare the behavior of a the five-cell cavity operating in
$\pi$-mode to a single-cell cavity. In the absence of any imperfections, we
describe~\cite{VZAP} the voltage $V$ in a single-cell resonator that is driven
with current $I$ by
\begin{equation}
  \frac{dV}{dt}=-\oo V +\frac{\oo R}{1+\beta} I
  \qquad\mathrm{with}\qquad
  \oo=\frac{\ho}{2 Q_L}=(1+\beta)\frac{\ho}{2Q}\ .
\end{equation}
This equation straightforward to integrate such that the the voltage after starting
up the generator is given by
\begin{equation}
  V(t) = \left( 1-e^{-\oo t}\right) V_{\infty}
  \qquad\mathrm{with}\qquad
  V_{\infty}=\frac{IR}{1+\beta}\ ,
\end{equation}
where $V_{\infty}$ is the asymptotic steady-state voltage. After the generator has
turned off, the voltage decreases exponentially $V(t)=V_ae^{-\oo t}$, where $V_a$
is the voltage immediately before the generator is turned off.
\par
Figure~\ref{fig:comp} shows the voltage in the single-cell cavity superimposed on the
voltages in cells~1, 3, and 5 of a five-cell cavity operated in $\pi$-mode.
The latter was already shown in the upper plot in Figure~\ref{fig:fiveb}. In the
single-cell model, we have to increase $Q$ and thereby the time constant $1/\oo$ by
a factor of five to account for filling five cells, rather than a single cell. But
this leads to almost perfect agreement of the curves in Figure~\ref{fig:comp} and
largely justifies the commonly-used approximation of a multi-cell cavity by a single-cell
equivalent model.
\par
After considering the perfect $\pi$-mode cavity, we now turn to the consequences
of imperfections such as incorrect coupling $\kappa_i$ or deviations of one cell's
unloaded $Q$-value $Q_i$ and resonance frequency $\ho_i$ from the design values.
\section{Incorrect coupling $\kappa_i$ between cells}
\label{sec:kappa}
We now assume that the $\kappa_i$ possibly differ from their design value value
$\kappa$ by $\Dk_i$. We also assume that all capacitances $C_i$ are equal and
that leads to $\kappa'_j=\kappa_{j-1}$. Furthermore, the coupling to the beam pipe
can differ from the design value according to $\kappa_a=2\kappa+\Delta\kappa_a$
and $\kappa_b=2\kappa+\Delta\kappa_b$. This causes us to modify Equation~\ref{eq:c1t}
and~\ref{eq:c1tp}
for the first cell as follows
\begin{eqnarray}\label{eq:c1tk}
  &&\left(2+6\kappa+\kappa_1+\Dk_a\right)\frac{d\tV_1}{dt} -\kappa_1\frac{d\tV_2}{dt}\\
  &&\qquad=-\frac{\ho}{Q'_L}\tV_1
         +i\frac{\ho}{\sqrt{1+4\kappa}}\left[(2\kappa-\kappa_1-\Dk_a)\tV_1+\kappa_1\tV_2\right]
         + \frac{\ho R_1}{Q_1}\frac{\tilde I_g}{n}\ .\nonumber
\end{eqnarray}
For the cells in the middle of the cavity Equations~\ref{eq:cjt} and~\ref{eq:cjtp}
are modified to
\begin{eqnarray}\label{eq:cjtk}
  &&-\kappa_{j-1} \frac{d\tV_{j-1}}{dt}
     +\left(2+4\kappa+\kappa_{j-1}+\kappa_j\right)\frac{d\tV_j}{dt}
  -\kappa_j \frac{d\tV_{j+1}}{dt}\\
   &&\qquad\quad = -\frac{\ho}{Q}\tV_j+\frac{i\ho}{\sqrt{1+4\kappa}}\left[\kappa_{j-1}\tV_{j-1}
      +\left(4\kappa-\kappa_{j-1}-\kappa_j\right)\tV_j+\kappa_j\tV_{j+1} \right]\ .\nonumber
\end{eqnarray}
Finally Equations~\ref{eq:cnt} and~\ref{eq:cntp}, describing the last cell~$n$, become
\begin{eqnarray}\label{eq:cntk}
  &&-\kappa_{n-1} \frac{d\tV_{n-1}}{dt}
  +\left(2+6\kappa+\kappa_{n-1}+\Dk_b\right)\frac{d\tV_j}{dt} \\
   &&\qquad\qquad = -\frac{\ho}{Q}\tV_n +\frac{i\ho}{\sqrt{1+4\kappa}}\left[\kappa_{n-1}\tV_{n-1}
      +\left(2\kappa-\kappa_{n-1}-\Dk_b\right)\tV_n\right]\ .\nonumber
\end{eqnarray}
For the simulations we convert these equations to matrix-valued equations. Compared
to Equation~\ref{eq:dynpaux}, only the matrices $D$ and $B$ need to be modified to
{\tiny
\begin{eqnarray}
  D&=&\left(\begin{array}{ccccc}
     2+6\kappa+\kappa_1+\Dk_a & -\kappa_1 & 0 & 0 & 0 \\
     -\kappa_1 & 2+4\kappa+\kappa_1+\kappa_2 & -\kappa_2 & 0 & 0 \\
     0 & -\kappa_2 & 2+4\kappa+\kappa_2+\kappa_3 & - \kappa_3 & 0 \\
     0 & 0 & -\kappa_3 & 2+4\kappa+\kappa_3+\kappa_4 & - \kappa_4 \\
     0 & 0 &  0 &-\kappa_4 & 2+6\kappa+\kappa_4+\Dk_b
   \end{array}\right)\nonumber\\
  B&=&\frac{\ho\dt}{\sqrt{1+4\kappa}}\left(\begin{array}{ccccc}
              2\kappa-\kappa_1-\Dk_a & \kappa_1 & 0 & 0 & 0\\
              \kappa_1 & 4\kappa-\kappa_1-\kappa_2 & \kappa_2 & 0 & 0\\
              0 & \kappa_2 & 4\kappa-\kappa_2-\kappa_3 & \kappa_3 & 0\\
              0 & 0 & \kappa_3 & 4\kappa-\kappa_3-\kappa_4 & \kappa_4\\
              0 & 0 & 0 & \kappa_4 & 2\kappa-\kappa_4-\Dk_b
       \end{array}\right)\ .
\end{eqnarray} }
It is easy to see that these expressions revert to those in Equation~\ref{eq:dynpaux}
if all $\kappa_i$ are equal to $\kappa$ and both $\Dk_a$ and $\Dk_b$ are zero.
\par
\begin{figure}[tb]
\begin{center}
\includegraphics[width=0.57\textwidth]{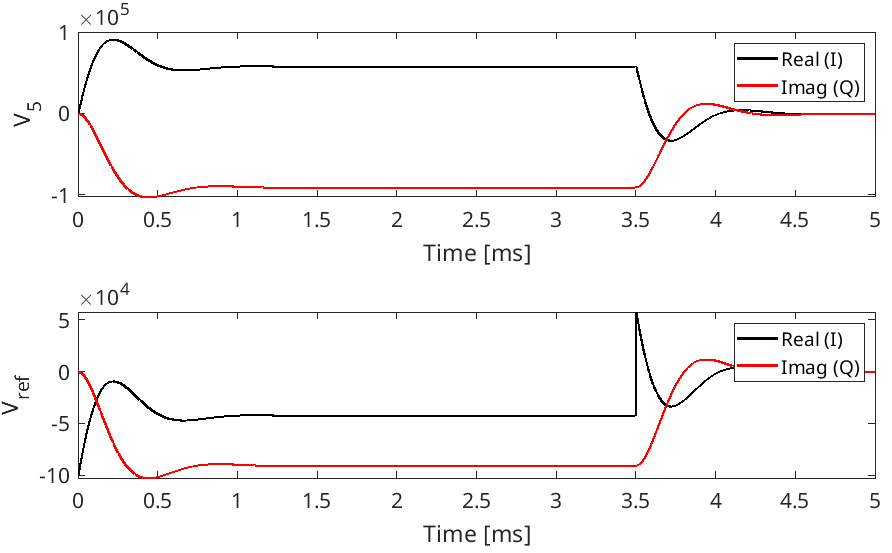}
\includegraphics[width=0.4\textwidth]{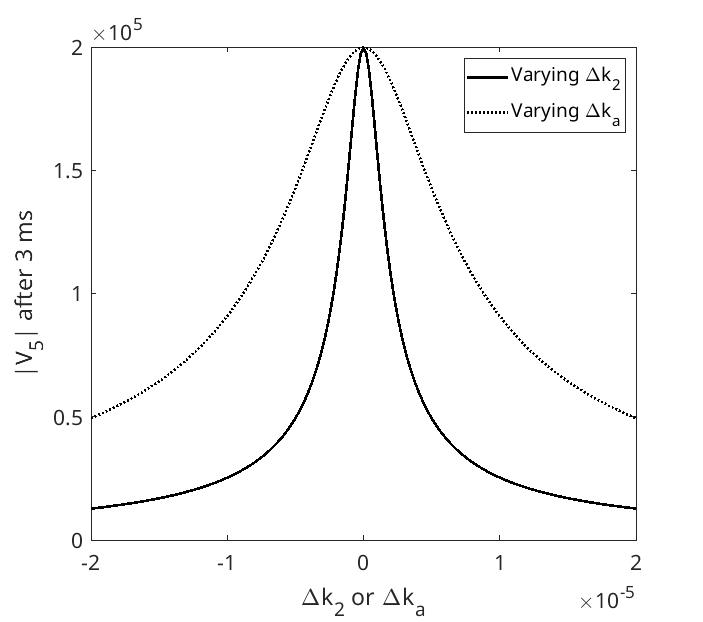}
\end{center}
\caption{\label{fig:fivedk2}The left plot corresponds to Figure~\ref{fig:fivec},
  only here the coupling between the second and third cell is increased by
  $\Dk_2=2\times 10^{-6}$ from the design value of $\kappa_2=0.01$. The right
  shows the amplitude of the voltage in the fifth cell $|V_5|$ after 3\,ms
  versus $\Dk_2$ (solid) and $\Dk_a$ (dotted) in the range $\pm2\times 10^{-5}$.}
\end{figure}
The left-hand plot in Figure~\ref{fig:fivedk2} shows the I and Q phase
of the field in the fifth cell on the top and the reflected signals on
the bottom. Here the coupling between the second and third cell is
increased by $\Dk_2=2\times10^{-6}$ compared to the situation shown in
Figure~\ref{fig:fivec}. On the top we observe that the maximum achievable
signal is substantially reduced from $2\times 10^5$ to $1.3\times 10^5$
with respect to Figure~\ref{fig:fivec} and in both top and  bottom plot
a significant imaginary phase (Q-phase, shown in red) appears. Moreover,
some ringing in the signal is visible. It turns out that the eigenvalue
of the $\pi$-mode, which was purely real in the situation depicted in
Figure~\ref{fig:fivec}, now acquires a small imaginary component, which
causes the onset of an oscillation. Increasing $\Dk_2$ further makes the
ringing very obvious, though we do not show any plots.
\par
The solid line in the plot on the right-hand side in Figure~\ref{fig:fivedk2}
shows the amplitude of the field in the fifth cell $|V_5|$ as a function of
$\Dk_2$, which displays a distinct resonance-like behavior. The
full-width-at-half-maximum
of the curve is $4.2\times 10^{-6}$, which may serve as a specification
for the tolerance for manufacturing the cavity and especially the cell-to-cell
coupling. The dotted line shows the corresponding response of varying the
coupling to the beam pipe $\kappa_a$ by $\Dk_a$ in the same range. We find
that the curve is about three times as wide, indicating slightly relaxed
tolerances for manufacturing the cavity ends.
\par
In the next section, we perform a comparable analysis for variations
in the unloaded quality factor $Q_i$ of the cells.
\section{Incorrect quality factor $Q_i$ of cells}
\label{sec:Q}
\begin{figure}[tb]
\begin{center}
\includegraphics[width=0.6\textwidth]{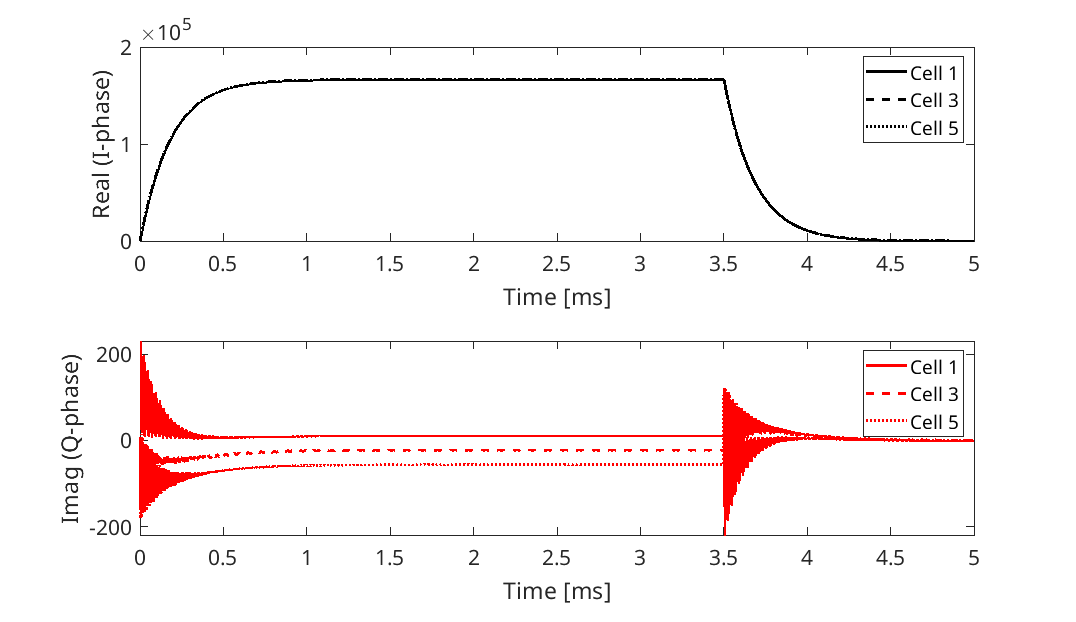}
\includegraphics[width=0.38\textwidth]{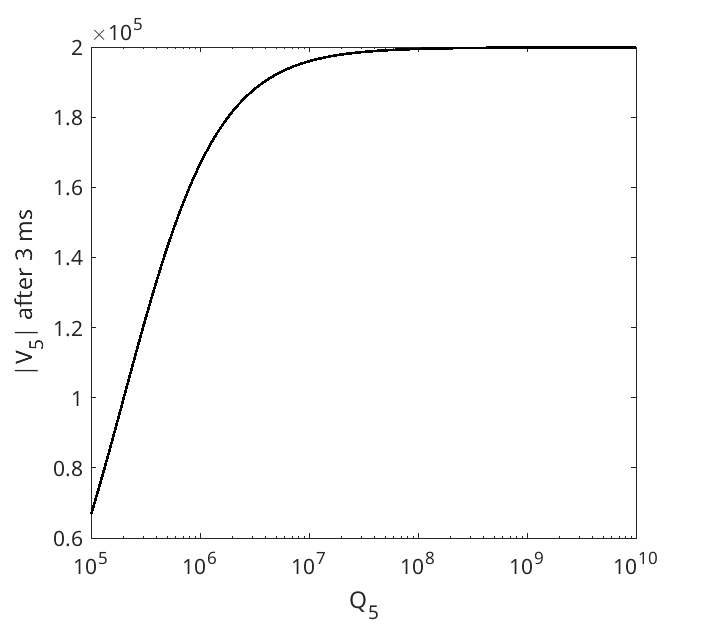}
\end{center}
\caption{\label{fig:fiveQ}The left plot corresponds to Figure~\ref{fig:fiveb},
  only here the quality factor in the fifth cell $Q_5$ is reduced by a factor
  1000 to $10^6$. The plot on the right shows the voltage amplitude in the
  fifth cell $|V_5|$ after 3\,ms as a function of $Q_5$ in the range $10^5$ to $10^{10}$.}
\end{figure}
Losses in the cells are modeled by varying the cell's resistance $R_i$, which
predominantly affects the quality factor $Q_i=R_i\sqrt{C_i/L_i}$, but also the
coupling $\beta=R_1/n^2Z_0$ of the first cell. To account for the losses, we
only have to modify the array $A$ in Equation~\ref{eq:dynpaux} to
\begin{equation}
  A=\diag\left((1+\beta)\frac{\ho\dt}{Q_1}, \frac{\ho\dt}{Q_2}, \frac{\ho\dt}{Q_3},
    \frac{\ho\dt}{Q_4},\frac{\ho\dt}{Q_5}\right)
\end{equation}
and leave the other quantities unaffected. The plot on the left-hand side in
Figure~\ref{fig:fiveQ} corresponds to Figure~\ref{fig:fiveb}, which shows the
unperturbed system. These figures show the I and Q-phases of the voltages
in cells 1, 3, and 5 during a pulse. Only here the quality
factor of the fifth cell $Q_5$ is reduced by a factor 1000 from $10^9$ to
$10^6$. The upper graph shows the real (I-phase) part of the voltages. Here
we observe a small reduction of the amplitude compared to Figure~\ref{fig:fiveb}.
The reduction of $Q_5$ causes the lower graph, showing the imaginary (Q-phase),
to exhibit a small splitting of this phase in the three cells. We interpret
this as the fifth cell absorbing most of the power that has to pass through
the preceding cells and the losses are somewhat bigger than can be replenished
continuously by the generator.
\par
The plot on the right-hand side in Figure~\ref{fig:fiveQ} shows the reduction
of the  absolute value of the voltage in the fifth cell $|V_5|$ as a function
of $Q_5$. As long as only $Q_5$ is reduced and the dissipated energy does not
quench neighboring cells, the voltage $|V_5|$ is only weakly affected, unless
it is reduced below a few times $10^6$.
\par
\begin{figure}[tb]
\begin{center}
\includegraphics[width=0.47\textwidth]{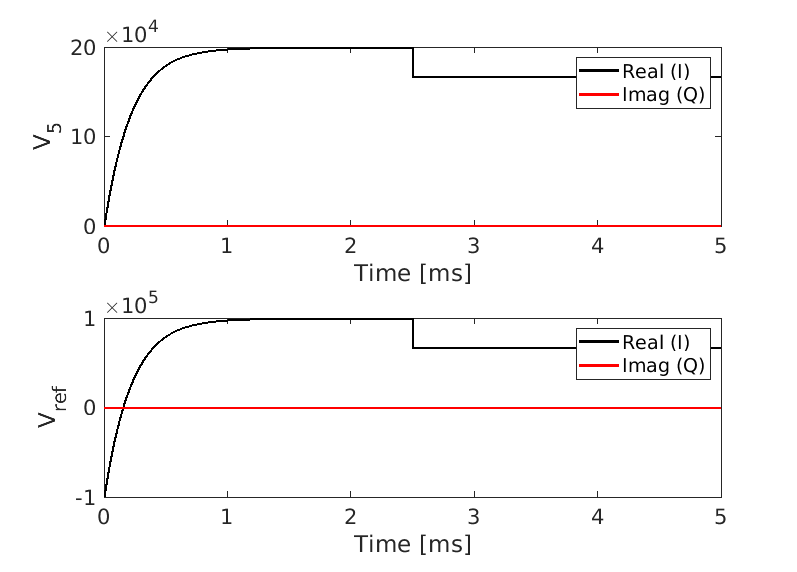}
\includegraphics[width=0.47\textwidth]{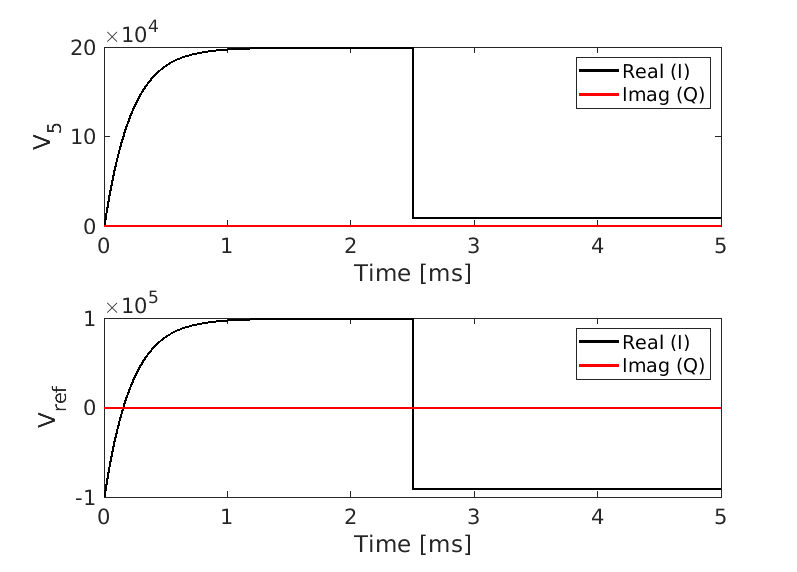}
\includegraphics[width=0.47\textwidth]{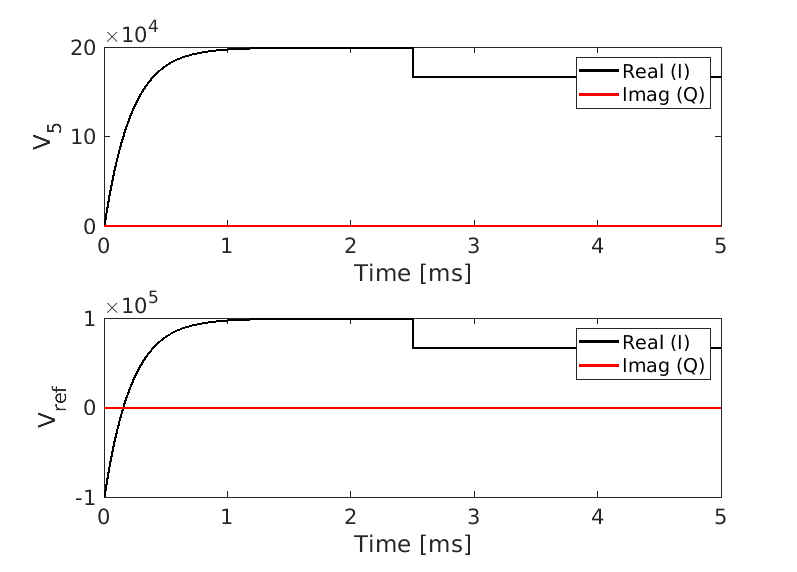}
\includegraphics[width=0.47\textwidth]{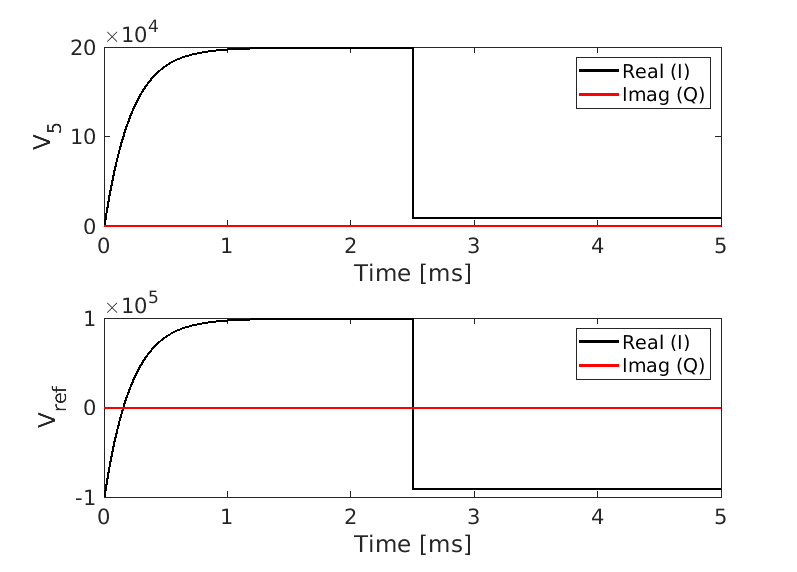}
\end{center}
\caption{\label{fig:quench}The field in the fifth cell $V_5$ and the reflected
  signal $V_{ref}$ while the first cell quenches after 2.5\,ms. The resistance
  $R_1$ is decreased by $x=10^3$ on the left-hand plot and by $10^5$ on the
  right-hand plot.}
\end{figure}
Increased losses in the first cell need a little extra attention, because both
$Q_1$ and $\beta$ are affected. Let us assume that the resistance $R_1$ is decreased
by some factor $x$. A big quench is characterized by a large value of $x$, say $10^5$.
This factor reduces $Q_1$ to $Q_1/x$ and $\beta$ to $\beta/x$, such that the first
entry in the matrix $A$ will become $(1+\beta/x)/(Q_1/x)=(x+\beta)/Q_1$ where $Q_1$
and $\beta$ are the default values. Here we see that as long as $x$ is smaller than
$\beta$ a quench in the first cell will only have a modest effect on the time
constant of the first cell, which is the first entry in the matrix $A$.
\par
Each of the plots in Figure~\ref{fig:quench} shows the field in the fifth cell~$V_5$
and the reflected signal $V_{ref}$ in situations where a quench happens after 2.5\,ms.
In the plots on the top row the first cell quenches. We simulate it by decreasing the
resistance $R_1$ by a factor $x=10^3$ (left) and $x=10^5$ (right). The weaker
quench, shown on the left, only leads to a moderate drop of the voltage $V_5$,
whereas the bigger quench, shown on the right, leads to  an almost complete collapse
of the field in the cavity and corresponding change in the reflected signal. The
two plots in the bottom row correspond to quenches with equal magnitude of the
fourth cell, that we model by reducing $R_4$. We find that the behavior of the
signals $V_5$ and $V_{ref}$ is very similar to that shown in the top row.
\par
Finally, let us consider what happens if one or several cells are detuned.
\section{Detuned cells}
\label{sec:omega}
In this section we assume that the resonance frequency $\ho_j$ of cell~$j$
deviates from the design value $\ho$ by $\Do_j=\ho_j-\ho$. Moreover, for the
$\pi$-mode we have $\ho^2/\omega^2_{\pi}=1+4\kappa$ which gives us
\begin{equation}
  \frac{\ho^2_j}{\omega_{\pi}^2}=\frac{\ho^2+2\ho\Do_j+\Do_j^2}{\omega_{\pi}^2}
  \approx 1+4\kappa+2\frac{\Do_j}{\ho}
  \approx 1+4\kappa + \delta_j
\end{equation}
with the abbreviation $\delta_j=2\Do_j/\ho$ to describe the relative
detuning of cell~$j$.
\par
Assuming all other parameters to be equal their design values, the equation for
the first cell becomes
\begin{eqnarray}\label{eq:c1td}
&&  \left(2+7\kappa+\delta_1\right)\frac{d\tV_1}{dt} -\kappa\frac{d\tV_2}{dt}\\
&&\qquad\qquad  =-\frac{\ho}{Q'_L}\tV_1
  +i\frac{\ho}{\sqrt{1+4\kappa}}\left[(\kappa+\delta_1)\tV_1+\kappa\tV_2\right]
  + \frac{\ho R_1}{Q_1}\frac{\tilde I_g}{n}\ .\nonumber
\end{eqnarray}
For cell~$j$ in the middle of the cavity we obtain
\begin{eqnarray}\label{eq:cjtd}
&&  -\kappa \frac{d\tV_{j-1}}{dt} +\left(2+6\kappa+\delta_j\right)\frac{d\tV_j}{dt}
     -\kappa \frac{d\tV_{j+1}}{dt}\\
  &&\qquad\qquad = -\frac{\ho}{Q}\tV_j+i\frac{\ho}{\sqrt{1+4\kappa}}
     \left[\kappa\tV_{j-1} +(2\kappa+\delta_j)\tV_j+\kappa\tV_{j+1} \right]
     \nonumber
\end{eqnarray}
and the equation for the last cell is
\begin{eqnarray}\label{eq:cntd}
  && -\kappa \frac{d\tV_{n-1}}{dt}+\left(2+7\kappa+\delta_n\right)\frac{d\tV_j}{dt}\\
  &&\qquad\qquad = -\frac{\ho}{Q}\tV_n +i\frac{\ho}{\sqrt{1+4\kappa}}
      \left[\kappa\tV_{n-1}+(\kappa+\delta_n)\tV_n\right]\ .\nonumber
\end{eqnarray}
In all equations we approximated $\ho_j/Q_j$ by $\ho/Q$ and thus neglected the 
very small effect of the detuning on damping. This leaves the matrix $A$, used in
the simulations, unaffected. On the other hand, the change of a cell's resonance 
frequency mainly affects the matrices $D$ and $B$ 
\begin{eqnarray}
  D&=&\left(\begin{array}{ccccc}
     2+7\kappa+\delta_1 & -\kappa & 0 & 0 & 0 \\
     -\kappa & 2+6\kappa+\delta_2 & - \kappa & 0 & 0 \\
     0 & -\kappa & 2+6\kappa+\delta_3 & - \kappa & 0 \\
     0 & 0 & -\kappa & 2+6\kappa+\delta_4 & - \kappa \\
     0 & 0 &  0 &-\kappa & 2+7\kappa+\delta_5
   \end{array}\right)\nonumber\\
  B&=&\frac{\ho\dt}{\sqrt{1+4\kappa}}\left(\begin{array}{ccccc}
       \kappa+\delta_1& \kappa & 0 & 0 & 0\\
              \kappa & 2\kappa+\delta_2 & \kappa & 0 & 0\\
              0 & \kappa & 2\kappa+\delta_3 & \kappa & 0\\
              0 & 0 & \kappa & 2\kappa+\delta_4 & \kappa\\
              0 & 0 & 0 & \kappa & \kappa+\delta_5
 \end{array}\right)\ 
\end{eqnarray}
which makes it straightforward to incorporate detuned cells in the simulations.
\par
\begin{figure}[tb]
\begin{center}
  \includegraphics[width=0.65 \textwidth]{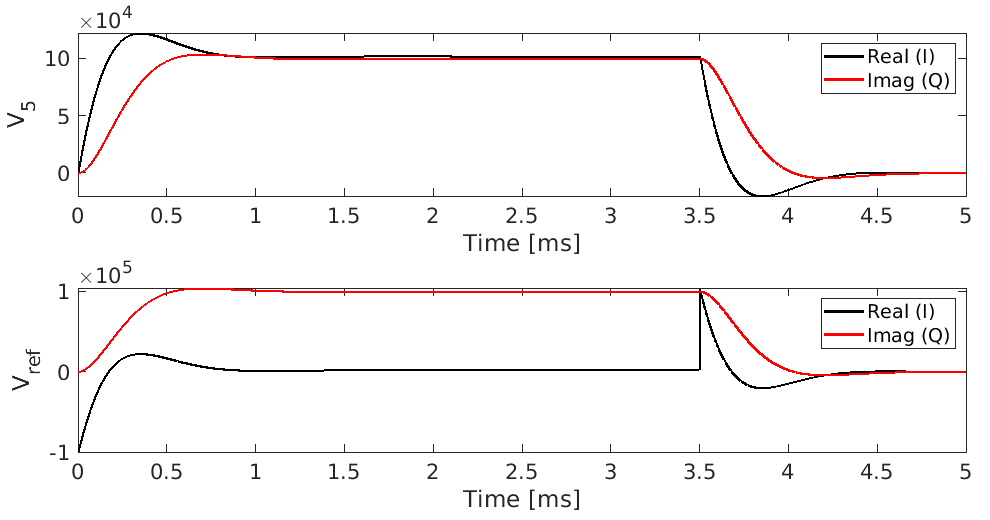}
  \includegraphics[width=0.34\textwidth]{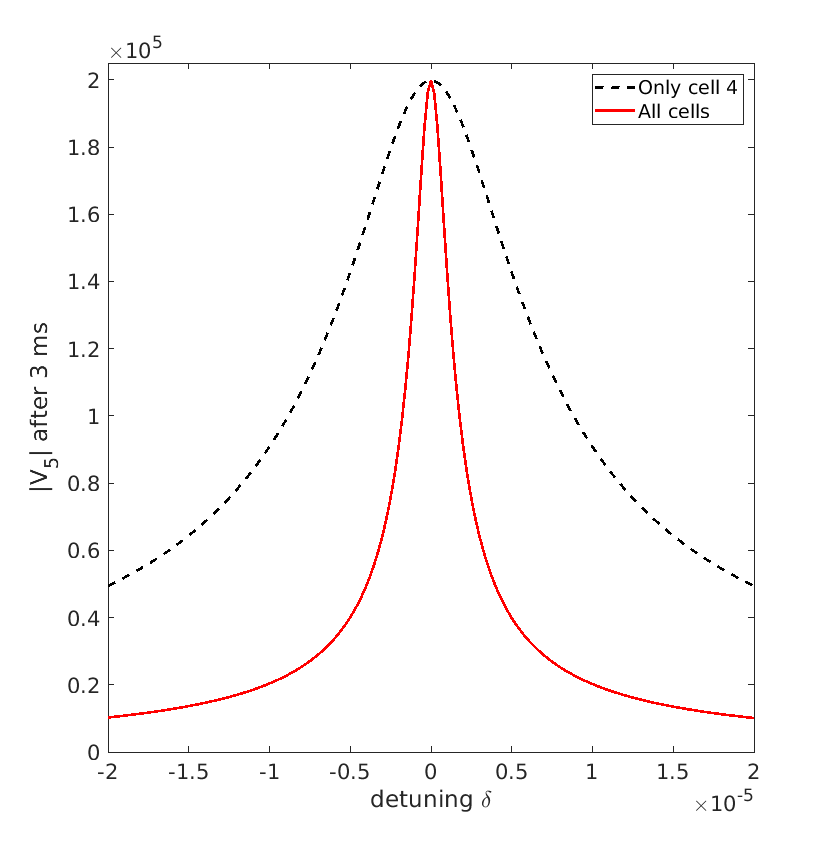}
\end{center}
\caption{\label{fig:fivedetuning}This figure corresponds to Figure~\ref{fig:fiveQ},
  but on the left-hand plot the fourth cell is detuned by $\delta_4=5\times 10^{-6}$. 
  The right-hand plot shows the voltage amplitude in the fifth cell $|V_5|$ plotted 
  against the detuning $\delta$ of the fourth cell (black dashes) and if all cells
are detuned by the same amount (red).}
\end{figure}
In the simulations in this section we keep all parameters at their default values, only
the detuning is varied. The left-hand plot in Figure~\ref{fig:fivedetuning} shows 
the voltage in the fifth cell $V_5$ in the upper panel and the reflected signal $V_{ref}$
on the lower panel. In this simulation, the fourth cell is detuned by $\delta_4=5\times10^{-6}$ 
which approximately corresponds to twice the loaded bandwidth $(1+\beta)/2Q$ of the cavity. 
We see that the real and imaginary phase of $V_5$ and $V_{ref}$ have similar magnitude, 
indicating the substantial influence of the detuning. We also observe that the amplitude 
of $V_5$ is reduced. In passing we note that detuning all cells by $\delta=10^{-6}$,
which is one fifth of $\delta_4$, the plot of $V_5$ and $V_{ref}$ is virtually unchanged.
In other words, from observing the transmitted $V_5$ and the reflected signal $V_{ref}$
we cannot tell whether one cell is detuned or all are detuned by  a smaller amount.
Finally, the right-hand plot in Figure~\ref{fig:fivedetuning} shows the amplitude in
the fifth cell $|V_5|$ versus the detuning. The dashed black curve corresponds to the
fourth cell detuned and the red curve corresponds to all cells detuned by the same
amount~$\delta$. We see that the red curve is much narrower than the black one,
indicating the much higher sensitivity when detuning all cells, compared to detuning
only a single cell.
\section{Conclusions}
We found that multi-cell cavities behave very similar to single-cell cavities,
as is witnessed by the close agreement of the two curves in Figure~\ref{fig:comp}.
The reason is that it takes only on the order of microseconds for the $\pi$-mode
to form, which is shown in Figure~\ref{fig:fivea}. The other modes are excited
but they do not absorb power to build up a coherent mode. Instead, they damp
away as shown in the upper four plots in Figure~\ref{fig:modamp}, where the
individual time constants are related to the real part of the eigenvalues shown
in Figure~\ref{fig:eigval}. Basically, apart from short-lived transients of the
other modes, the $\pi$-mode dominates the dynamics of the electro-magnetic
fields in multi-cell cavities and oscillates as an entity, rather than the
weakly-coupled fields in the cells oscillating independently.
\par
That the $\pi$-mode oscillates as an entity has ramifications for the tolerances
to which the cells must be manufactured, because the $\pi$-mode averages over
the contributions of the individual imperfections. We determined the tolerance
with respect to the cell-to-cell coupling $\kappa$ and show the result on the
right-hand plot in Figure~\ref{fig:fivedk2}, which shows a resonance-like curve
whose width determines the sensitivity with respect to $\kappa$. Furthermore,
the right-hand plot in Figure~\ref{fig:fivedetuning} shows the sensitivity
with respect to detuning of individual cells, which is more relaxed than detuning
all cells by the same amount. In Figure~\ref{fig:quench} we found that a
quench in a single cell can be stabilized if it is ``small'' in the sense
that the losses are comparable to the losses that escape through the input
coupler, which is given by $\beta$. If the reduction $x=Q/Q_{quench}\gg\beta$
of one cell, the quench leads to a collapse of the $\pi$-mode as shown on
the right-hand plots in Figure~\ref{fig:quench}.
\par
The methods developed for this report can be easily extended to other cavities
and can then be used to estimate manufacturing tolerances for a number of
imperfections, such as evenness of $Q$-values or statistically distributed
couplings $\kappa$ or detuning $\delta=2\Do/\ho$.
\par
Of particular interest are the initial high-frequency oscillations shown in
Figure~\ref{fig:fivea}. Measuring them experimentally with a high-speed
data-acquisition system and theoretically determining how they are related
to cavity imperfections using methods developed for measuring $Q_0$~\cite{Q0}
seems to be a worthwhile enterprise. 
\par
This work was produced in part by Jefferson Science Associates, LLC under
Contract No. AC05-06OR23177 with the U.S. Department of Energy. Publisher
acknowledges the U.S. Government license and provide public access under
the DOE Public Access Plan (\url{http://energy.gov/downloads/doe-public-access-plan}).
%
%
\bibliographystyle{plain}

\begin{thebibliography}{M}
  %
\bibitem{SCHILCHER}
  T. Schilcher, {\em Vector sum control of pulsed accelerating fields in lorentz
    force detuned superconducting cavities,} Dissertation, Universit\"at Hamburg, 1998.
\bibitem{SYSID}
  V. Ziemann, {\em Simulations of real-time system identification of superconducting cavities
    with a recursive least-squares algorithm,} Physical Review Accelerators and Beams 26 (2023)
  112003.
\bibitem{PKH}
  H. Padamsee, J. Knobloch, T. Hays, {\em RF Superconductivity for Accelerators, 2nd ed.,}
  Wiley-VCH, Weinheim, 2008.
\bibitem{LIEPE}
  M. Liepe, {\em Superconducting Multicell Cavities for Linear Colliders,} Dissertation, 
  Universit\"at Hamburg, 2001.
\bibitem{PLASMA}
  T. Powers, N. Brock, T. Ganey {\em In Situ Plasma Processing of Superconducting
    Cavities at JLab, 2023 Update,} Proceedings of the 21st International Conference
    on RF Superconductivity SRF2023 (2023) 701.
\bibitem{VZAP}
  V. Ziemann, {\em Hands-On Accelerator Physics Using MATLAB, 2nd ed.,}
  CRC Press, Boca Raton, 2025.
\bibitem{Q0}
  V. Ziemann, {\em A new method to measure the unloaded quality factor of superconducting cavities,}
  Physical Review Accelerators and Beams 27 (2024) 032001;
%
\end{thebibliography}

\end{document}